\documentclass[12pt,preprint]{aastex}
\usepackage{amsmath,longtable,natbib}
\shorttitle{binary pulsars in GCs} \shortauthors{Bagchi \& Ray}
\begin{document}

\title{Radio pulsar binaries in globular clusters: their orbital eccentricities
and stellar interactions}

\author{Manjari Bagchi and Alak Ray}
\affil{Tata Institute of Fundamental Research, Mumbai 400005, India}

\begin{abstract}

High sensitivity searches of globular clusters (GC) for radio pulsars by improved
pulsar search algorithms and sustained pulsar timing observations
have so far yielded
some 140 pulsars in more
than two dozen GCs. The observed distribution of orbital eccentricity and period of binary
radio pulsars in GCs have imprints of
the past interaction between single pulsars and binary systems or of binary pulsars and 
single passing non-compact stars. It is seen that GCs have different groups of pulsars.
These may have arisen out of exchange or merger of a component of the binary with the 
incoming star or a ``fly-by" in which the original binary remains intact but undergoes a 
change of eccentricity and orbital period. We consider the genesis of the
distribution of pulsars using analytical and computational tools
such as STARLAB, which performs numerical scattering experiments
with direct N-body integration. Cluster pulsars with intermediate eccentricities can 
mostly be accounted for by fly-bys whereas those with high eccentricities are likely 
to be the result of exchanges and/or mergers of single stars with the binary companion of 
the pulsar, although there are a few objects which do not easily fit into this description.
The corresponding distribution for galactic field pulsars shows notable differences from the 
GC pulsar orbital period and eccentricity distribution. The long orbital period pulsars in 
the galactic field with frozen out low eccentricities are largely missing from the globular 
clusters, and we show that ionization of these systems in GCs cannot alone account for the 
peculiarities.

\end{abstract}

\vskip 0.6 cm

\keywords{ pulsars: general --- globular clusters: general --- stellar dynamics --- methods: N-body simulations --- binaries (including multiple): close}

\section{Introduction }
\label{sec:intro}

Globular clusters (GCs) are dense spherical collection of stars orbiting around the galactic
center containing a significant number of binary stars. The first millisecond radio pulsar was
discovered in a GC M28 only in 1987 \cite{lyn87} and GCs in general
were regarded as spun-up pulsar nurseries \cite{alp82}. How old and evolved stellar systems
like GCs can contain apparently young, active objects like pulsars have led to suggestions,
based on lifetime constraints from orbital decay by gravitational radiation \cite{pri91},
that
some of these systems have formed recently, or are forming even today.
Interaction between
binary stars and single stars in GCs is believed to be an important dynamical process as
the resulting binary systems
may provide a substantial source of energy for the GCs,
as the binding energies of a few, very close binaries (like neutron star binaries) can
approach that of a moderately massive host GC \cite{spi87,hut03}.
Although the importance there is due to exchange of energy, interesting constraints and
conclusions can be derived about lifetimes of binaries from the rate at which their orbital
eccentricity is changed by encounter with other stars. A key aspect here
is that the orbital eccentricities of binary millisecond radio pulsars can be measured with
extraordinary precision.

Pulsar spin-up and recycling are thought to be enhanced
in globular clusters by binary-single star interactions. This is because binary interactions can
increase the cross section for recycling processes and weaken the stellar density dependence
of the probability of a neutron star without visible radio pulses evolving into a reborn
pulsar after the (binary) interaction \cite{sig93}. After the interaction
some of the pulsars may be ejected from the core of the cluster or even from the GC altogether,
and their distributions of orbital properties can be
predicted in terms of the interaction model.
Thus, observable parameters of the neutron stars
and their binary companions in globular clusters, such as spin,
orbital period and eccentricity, projected radial position in the
cluster, companion mass and their distributions provide a tracer of the
past history of dynamical interactions of the binary NSs in individual GCs.
These parameters can even provide a valuable test-bed to examine the theoretical scenarios of
formation and evolution of recycled pulsars.
Among the different types of neutron star binaries, binary radio pulsars can be timed more easily
and accurately by ground-based telescopes over well-separated epochs, as the
underlying neutron stars are less prone to noise
than in the case of X-ray pulsars which have
episodically varying accretion torques.
These lead to the easier measurement of orbital parameters.
Eccentric millisecond pulsar binaries can also be important probes of neutron star physics
since they provide a way to constrain the masses of the fully-recycled pulsars. Pulsar
timing can measure the orbital advance of the periastron of the elliptical orbit; this advance
is dominated by general relativistic effects if the companion star is a compact one, and
can determine the total mass of the system. The recycling scenario creates binary millisecond
pulsars in circular orbits as during the accretion phase there is a strong tidal coupling leading
to orbit circularization. Therefore, the eccentric systems of fully recycled (millisecond period)
pulsar binaries found in globular clusters are produced only through stellar interactions where
density of stars is high \cite{ras95}.

Till today, a total of
140 pulsars in 26 GCs\footnote{Information on these
pulsars are found from P. Freire's webpage updated August 2008,
http://www.naic.edu/$\sim$pfreire/GCpsr.html, compiled from radio
timing observations by many groups.} have been discovered. Among them 74 are known as binaries (in 23 GCs), 59 are known as
isolated and 7 have no published
timing/orbital solutions. In the present work, we concentrate on
binaries.  For one of them (PSR J2140$-$2310B in M30),  orbital parameters are
not well determined, only lower limit of $P_{orb}$ and $e$ are
known as $P_{orb} > 0.8$ days and $e > 0.52$. So we exclude it
from our analysis and the total number of pulsars in become 73. We study the distribution of these 73 pulsars in the orbital
eccentricity - period plane with analytical results and our numerical experiments of scattering of stars simulated by direct
N-body integration tools. The corresponding distribution
of some of these observables for neutron star binaries in the
galactic field e.g. in the orbital eccentricity and period plane
have some significant differences. We discuss below these
differences in the light of interaction of stars with binaries
already formed or in the process of formation inside globular
clusters, due to the high stellar density in their cores compared
to that of the galactic disk.

This article is in continuation of our preliminary work in this area \cite{bag09} and provides further details and extensions. The present paper is organized as follows: in section \ref{sec:data} we discuss the characteristics of the presently available binary radio pulsar data and how they occupy different regions of the orbital phase space. Here we include both GC pulsars and disk pulsars as the later class was not considered in the earlier paper \cite{bag09}. Note that in Fig 1, 2 and 5, we plot the GC and galactic disk pulsars irrespective of their spins. Our conclusions are not significantly affected if we restrict the sample to $P_s<30$ ms pulsars which were more likely to have undergone atleast mild recycling. Details of methodology and notations have been discussed in section \ref{sec:types_int} which were not provided in our earlier paper \cite{bag09}. Section 3 is further subdivided into \S~\ref{sec:fly_by} for fly-by and \S~\ref{sec:starlab} for exchange and merger interactions. We report the details of the STARLAB runs used to simulate binary-single stars scattering and provide outputs for some of those runs $e.g$. cross-sections of different interactions for different stellar parameters, eccentricity - orbital period distribution after exchange/merger interactions $etc$. Next, in section \ref{sec:ion} we have a discussion on the role of ionization in affecting the distribution of pulsar characteristics where we show that ionization at present
cannot be responsible for the lack of long period binaries in the observed GCs. In section \ref{sec:disc} we discuss our analytical and computational results in relation
to the available data and in section \ref{sec:conclu} we give our conclusions.

\section{Archival data on binary radio pulsars in globular clusters}
\label{sec:data}

\begin{figure}[h!]
\epsscale{1.0} \plotone{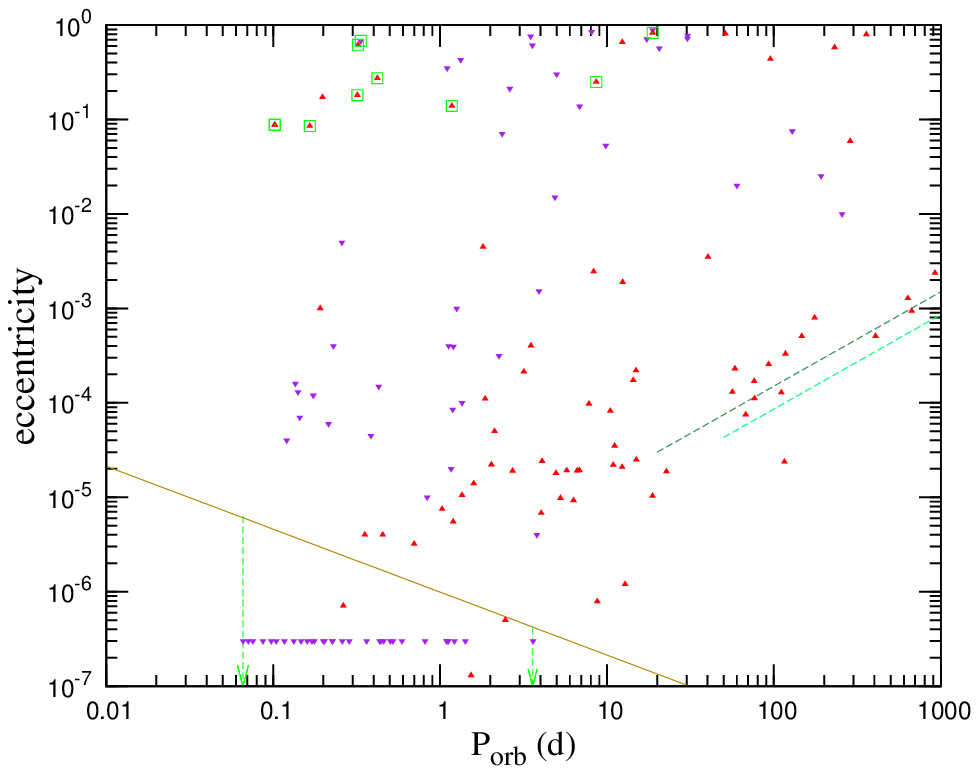} \caption{Plot of eccentricity vs
orbital periods for binary radio pulsars in and out globular
clusters. Purple $\blacktriangledown$s are for globular cluster
binaries and red $\blacktriangle$s are for disk binaries. The
solid line in the lower left corner is the line for ``limit of
timing sensitivity". The dashed lines in the right side represents
the eccentricity-orbital period relation as predicted by Phinney
(1992) - the upper line is the original line drawn by Phinney
(1992) and the lower line is using population II models. Double
neutron star binaries are enclosed by green $\square$s.
\label{fig:compare_gc_gal}}
\end{figure}

We summarize the relevant properties of the host GCs
\citep{web85} and binary pulsars therein in tables \ref{tb:gc_v_n}
and \ref{tab:psr_parms} respectively. In Fig
\ref{fig:compare_gc_gal}, we plot $e$ Vs $P_{orb}$ for all the
binary pulsars with known orbital solutions. Purple
$\blacktriangledown$s are for globular cluster binaries and red
$\blacktriangle$s are for disk binaries. A logarithmic scale in
both eccentricity and orbital periods are chosen, as the enormous
range of both variables and the regions occupied by observed
pulsars are less obvious in linear scales\footnote{There are
three disk binaries with $P_{orb} > 1000$ days such as (i) PSR
J0823+0159 with $P_{orb}=1232.40400$ days, $e=0.0118689$, (ii)
PSR J1302-6350 with $P_{orb}=1236.72404$ days, $e=0.8698869$,
(iii) PSR J1638-4725 with $P_{orb}=1940.9$ days, $e=0.9550000$.
These are excluded from our graph as we choose the range of
$P_{orb}$ up to 1000 days because our main concern, globular
cluster binaries  have always $P_{orb} < 1000$ days.}. 
Note that 10 GC binary pulsars have only upper limits to their eccentricities (see table \ref{tab:psr_parms}). In addition, one {\it galactic disk} binary pulsar PSR J1744-3922 ($P_{orb}=0.2$ days) also has only an upper limit to the eccentricity of $<0.001$.
Observed binaries in GCs can be categorized in three groups : (I) large
eccentricity pulsars ($e \geqslant 0.01$), 21 pulsars (22 if we
include M30B), (II) moderate eccentricity pulsars ($0.01 > e
\geqslant 2 \times 10^{-6}$), 20 pulsars and (III) small
eccentricity pulsars ($e \sim 0$), 32 pulsars. In the database,
orbital eccentricities of group (III) pulsars have been listed as
zero, and in this logarithmic plot we assign them an arbitrarily
small value of $e~=~3 \times 10^{-7}$. Actually for these
binaries, it is difficult to measure the eccentricities with
present day timing accuracy. This gives a limit to the minimum
eccentricity which can be measured as a function of orbital period
as given by Phinney (1992) :

\begin{equation}
\delta t = \left(e a / 2 c \right)sin~i \left[sin~\omega + sin (4
\pi t /P_{orb}-\omega) \right] \label{eq:timing_sensitivity1}
\end{equation}
giving
\begin{equation}
e_{min}= (\delta t) / (a sin~i /c) = \frac{4 \pi^2 c}{sin~i
\left[G(m_p+m_c)^{1/3}\right]}\frac{\delta t}{ P_{orb}^{2/3}}
\label{eq:timing_sensitivity2}
\end{equation}
where $i$ is the inclination angle of the binary, $\omega$ is the
angle between the line of nodes and the line of apsides, $m_p$ is
the pulsar mass, $m_c$ is the companion mass and $\delta t$ is
the timing accuracy. We call equation (\ref{eq:timing_sensitivity2})
as the ``limit of timing sensitivity" and is shown as the solid
line in the lower left corner of Fig \ref{fig:compare_gc_gal}
taking $m_p~=~1.4~M_{\odot}$, $m_c~=~0.33~M_{\odot}$,
$i~=60^{\circ}$, $\delta t~=~ 1 ~\mu{\rm sec}$. In addition we plot
on the right of Fig 1, dashed lines corresponding to the freeze-out
eccentricity - orbital period relation predicted by Phinney (1992) on the
basis of the fluctuation dissipation theorem for convective eddies in
the erstwhile red giant envelopes surrounding the white-dwarf cores
which end up as companions of the neutron stars. This relation is obtained
from the following expression:
\begin{equation}
\frac{1}{2} \mu \Omega_b^2 a ^2 <e^2>= 3.4 \times 10^{-5} (L^2
R_c^2 m_{env})^{1/3}
\end{equation}

\noindent where $\Omega_b~=~2\pi / P_b$ is the orbital angular
rotation frequency, $a$ is the semi-major axis of the binary, $e$
is the orbital eccentricity, L is the luminosity of the companion,
$R_c$ is the radius of the companion, $m_{env}$ is the envelop
mass of the companion, $m_p$ is the pulsar mass, $m_c$ is the
companion mass and $\mu~=(m_p m_c)/(m_p+m_c)$ is the reduced mass
of the system. Phinney used $R_c=0.125~a$ when the mass transfer
ceases and $e$ freezes. Then the above equation reduces to
$<e^2>^{1/2}= C_1~(P_{orb}/100)$ where $C_1$ is a function of
$T_{eff}$, $\mu$ and $m_{env}$. $P_{orb}$ is the orbital period
in days. Phinney (1992) calculated $C_1~ = ~1.5 \times 10^{-4}$
using average values of $m_{env}$ and $T_{eff}$ for population I
stars ($z~=~0.04$) \cite{ref70}. However for globular clusters,
population II stars are more appropriate. So we recalculated $C_1$ for
population II red giant stars which in general are smaller than
population I stars of similar mass and evolutionary stage, and
their binary orbital periods at Roche lobe overflow and
subsequent mass transfer are shorter than population I red giant
binaries \cite{web83}. We thus obtained $C_1=0.86 \times 10^{-4}$.
This line is plotted as the lower dashed green line in Fig.
\ref{fig:compare_gc_gal}.

The galactic field pulsars in the phase space of Fig. \ref{fig:compare_gc_gal} are also
shown here. Observational selection effects may be influencing the
distribution seen in this figure for the two sets \cite{cam05},
the most important selection effect being operative towards the
left of the diagram, namely, it is more difficult to detect
pulsars with larger DM and/or shorter periods, especially
millisecond pulsars in very short orbital period and highly
eccentric binaries; distance to the pulsars also is an important
selection effect, since at large distance only the brightest
pulsars can be observed). Nevertheless, it is clear that there is
a large abundance of pulsars with long orbital periods and
intermediate eccentricity near the eddy ``fluctuation dissipation"
lines among the {\it galactic disk} pulsars which are absent in
the GC pulsar population, where the important selection effects
are unlikely to play a major role. Other than this, the galactic
disk pulsars by and large seem to follow the above three
groupings for GC pulsars. The observed preponderance of double
neutron star binaries in the field pulsars among the very short
orbital period and large eccentricity orbits may be related to
selection effects (see above), but we note that there is at least
one DNS binary among the GC system: PSR J2127+1105C in M15 in
this group (the galactic field DNS binary pulsars are by and
large at small distances while M15 and other host GCs are usually
far away)\footnote{The issue of double neutron star binaries (DNS systems)
has been considered by Ivanova $et~al.$ (2008). The case for the
eccentricity distribution of galactic disk double neutron star
(DNS) binaries has been considered by Ihm, Kalogera \& Belczynski (2006). Since these
systems are the result of evolution of massive stars in binary systems,
if they formed directly in GCs, they would have formed early in
the life of a GC, since the lifetime of a massive star ($M \geq 8
M_{\odot}$) is of the order of only $10$ Myr. Most binaries that
have one component forming a neutron star are disrupted in the
process of the NS formation, especially those that are in low
mass systems, but those that survive tend to have progenitor
binaries that were rather tightly bound and subsequent
binary-single encounter mostly harden the NS binary. Ivanova $et~
al.$ (2008) find only 14 DNSs were formed dynamically during
the 11 Gyr for all the 70 models of GC population synthesis they
consider, most of these would have formed in Ter 5. The only
known case of a DNS in a GC is however that of M15 PSR
B2127+1105C. PSR J0514-4002A is a 5-ms pulsar is located in the
globular cluster NGC 1851; it belongs to
 a highly eccentric (e = 0.888) binary system, but the massive
companion here may be a white dwarf rather than a neutron star
(Freiere, Ransom \& Gupta 2008).
}.

We have also performed a cluster analysis test using the
statistical software ``{\it R}"\footnote{www.R-project.org} for
globular cluster binary pulsars. Here we assign random
eccentricities of the group III (very low eccentricity) pulsars
in such a way that they remain below the ``limit of timing
sensitivity" line. Performing a K-means test with 3 clusters, we
got 3 clusters of sizes 21 (high eccentricity), 18 (medium
eccentricity) and 34 (low eccentricity) with the sum of squares
from points to the assigned cluster centers as 19.92446, 10.21016
and 13.25632 respectively whereas the squares of inter-cluster
distances are as follows $d_{23}^2~=~7.568824$, $
d_{31}^2~=~30.50265$, $d_{12}^2~=~10.85488$. This indicates that
the three groupings referred to above are statistically separate
cluster of objects.

Formation scenario of binary millisecond pulsars suggests that
their eccentricity should be very small, $\sim 10^{-6}-10^{-3}$
\cite{phi92}. But there are many eccentric millisecond pulsars
inside globular clusters which are likely to be due to
interactions of the binary pulsars with single stars provided the
interaction timescale is less than the binary age. We take the
maximum age of the binaries in a GC to be the globular cluster
ages which are close to $\sim 10^{10}$ years. In subsequent
sections, we study different types of stellar interactions and
their effect on orbital eccentricities and periods. We include
where relevant, the effects of gravitational radiation on these
phase space variables.

\section{Orbital parameter changes due to stellar interactions
\& gravitational radiation}
\label{sec:types_int}

Globular clusters contain a large number of low mass X-ray binaries
compared to the galactic field; this  had led to the suggestion
\cite{fab75} that a binary is formed by tidal capture of a
non-compact star by a neutron star in the dense stellar
environment of the GC cores. If mass and
angular momentum transfer from the companion to the neutron star took place
in a stable manner, this could lead
to recycled pulsars in binaries or as single millisecond pulsars
(e.g. Alpar $et~al.$ 1982, Romani $et~al$ 1987, Verbunt  $et~al$
1987). However large energies may be deposited in tides
after tidal capture of a neutron star by a low mass main
sequence star in a GC. The resultant structural readjustments of the star in
response to the dissipation of the modes could be very
significant in stars with either convective or radiative damping
zones (Ray, Kembhavi \& Antia 1987; McMillan, McDermott \&
Taam 1987) and the companion star can undergo a size ``inflation"
due to its high tidal luminosity  which may exceed
that induced by nuclear reactions in the core, by a large factor. Efficiency of
viscous dissipation and orbit evolution is crucial to the
subsequent evolution of the system as viscosity regulates the
growth of oscillations and the degree to which the extended star is
bloated and shed. The evolution of the system could lead to
either a merger (leading to a Thorne Zytkow object) or a neutron
star surrounded by a massive disk comprising of the stellar
debris of its erstwhile companion or, in other cases, a low mass
X-ray binary or a even detached binary following envelope
ejection. Tidal oscillation energy can also be {\it returned } to
the orbit (i.e. the tide orbit coupling is a dynamical effect)
thus affecting the orbit evolution or the extent of dissipative
heating in the less dense companion. The evolutionary transition
from initial tidal dynamics to any likely final, quiescent binary
system is thus regulated by viscosity. Tide orbit coupling can
even lead to chaotic evolution of the orbit in some cases
\cite{mar95} but in the presence of dissipation with non-liner
damping, the chaotic phase may last only for about 200 yr after
which the binary may undergo a periodic circularization and after
about 5 Myr finish circularizing \cite{mar96}. Of the stars that
do not directly lead to mergers, a roughly equal fraction of the
encounters lead to binaries that either become unbound as a
result of de-excitation or heating from other stars in the
vicinity or they are scattered into orbits with large pericenters
(compared to the size of the non-compact star) due to angular
momentum transfer from other stars \cite{koc92}. Thus Kochanek's
``intermediate pericenter" encounters can lead to widened orbit
(even 100 days orbital period in an eccentric orbit) in tidal
encounters involving main sequence stars and neutron stars, which
in the standard scenario are attributed to encounters between
giants and neutron stars. The complexities of the dynamics of
tidal capture and subsequent evolution are manifold and attempts
to numerically simulate collisions of neutron stars with red
giants have been made with 3D Smoothed Particle Hydrodynamics
(SPH) codes (e.g. Rasio \& Shapiro 1991, Davis, Benz \& Hills
1991). We note in this context that Podsiadlowski, Rappaport \&
Pfahl (2002) have stated ``it is not only premature to rule out
tidal capture as a formation scenario for LMXBs, but that LMXBs
in globular clusters with well determined orbital periods
actually provide observational evidence in its favor". The
different formation channels of pulsars in binaries in GCs and
the birthrate problems of millisecond pulsar binaries vs LMXBs
were considered in \cite{ray90,kul90}.

Until the early eighties there was no evidence of a substantial population of primordial
binaries in any globular clusters, and it was even thought that GCs are significantly
deficient in binaries compared to a younger galactic population \cite{hut92}. Theoretical modeling of GCs was often started off as if all stars were singles.
During the 1980s several observational techniques began to yield a rich population of
binary stars in GCs (see Bassa $et~al.$ 2008 , Yan \& Mateo 1994,  Pryor $et~al.$ 1989). When the local binary fraction is substantial, the single star - binary interaction
can exceed the encounter rate between single stars by a large factor \cite{sig93}. The existence of a significant population of primordial binary population
in GCs indicated that three body processes have to be accounted for in any dynamical
study of binaries involving compact stars. An encounter between a field star and a binary
may lead to a change of state of the latter, e.g.:
i) the original binary may undergo a change of eccentricity and orbital period but otherwise
remain intact -- a ``preservation" or a ``fly-by" process; ii) a member of the binary may be
exchanged with the incoming field star, forming a new binary -- an ``exchange" process;
iii) two of the stars may collide
and merge into a single object, and may or may not remain bound to the third star
- a ``merger" process; iv) three of the stars may collide
and merge into a single object - a ``triple merger" process;
v) all three stars are unbound from an orbit - an ``ionization" process.
The ionization is always a ``prompt" process, whereas the others can be
either prompt or ``resonant" process. In resonant processes, the stars
form a temporarily bound triple system, with a negative total energy but
which decays into an escaping star and a binary, typically after 10-100
crossing times.

The value of the binding energy of the binary and the velocity of the incoming star determine the type of interaction the binary will
encounter. The critical velocity $v_c$ of the incoming star, for which the total energy (kinetic plus potential) of the three body system
is zero, can be defined as
\begin{equation}
v_c^2=G \frac{m_1 m_2 (m_1+m_2+m_3)}{m_3(m_1+m_2)}\frac{1}{a_{in}}
\label{eq:vc}
\end{equation}
where $m_1$ and $m_2$ are the masses of the binary members, $m_3$ is the mass of the incoming star, $a_{in}$ is the semi-major
axis of the binary and G is the gravitational constant.

In case of a binary-single star interaction, semi major axis of the final binary ($a_{fin}$) is related to the semi major axis of the
initial binary ($a_{in}$) as follows
(Sigurdsson \& Phinney 1993):
\begin{equation}
a_{fin}=\frac{1}{1-\Delta}\frac{m_a m_b}{m_1 m_2} ~ a_{in}
\label{eq:afin_gen}
\end{equation}
where $m_1$ and $m_2$ are masses of the members of the initial
binary, $m_a$ and $m_b$ are masses of the members of the final
binary. $\Delta$ is the fractional change of binary binding
energy $i.e.$ $\Delta = \left(E_{in}-E_{fin}\right)/E_{in}$. The
binding energies of the initial and final binaries are
\begin{equation}
E_{in} = -G \frac{m_1 m_2}{2 a_{in}}, ~~~~ E_{fin} = -G \frac{m_a
m_b}{2 a_{fin}}
\end{equation}
For fly-by, $m_a = m_1$, $m_b = m_2$
giving
\begin{equation}
a_{fin}=\frac{1}{1-\Delta}~a_{in} \label{eq:afin_fly}
\end{equation}
For exchange, $m_a = m_1$, $m_b = m_3$
(star 2 is being replaced by star 3) giving

\begin{equation}
a_{fin}=\frac{1}{1-\Delta}\frac{m_1 m_3}{m_1 m_2} ~ a_{in}
=\frac{1}{1-\Delta}~\frac{m_3}{m_2}~ a_{in}\label{eq:afin_ex}
\end{equation}
For merger, $m_a = m_1$, $m_b = m_2 +
m_3 $ (star 2 is being merged with star 3) giving

\begin{equation}
a_{fin}=\frac{1}{1-\Delta}\frac{m_1 (m_2+m_3)}{m_1 m_2}~ a_{in}
=\frac{1}{1-\Delta}\frac{m_2+m_3}{m_2}~ a_{in}
\label{eq:afin_merg}
\end{equation}
The actual value of $\Delta$ is not known. Putting $\Delta~=~0$
simplifies equations (\ref{eq:afin_gen}), (\ref{eq:afin_fly}),
(\ref{eq:afin_ex}) and (\ref{eq:afin_merg}) giving $a_{fin} =
a_{in}$ for fly-by interactions. Similar expressions can be
derived for other cases $e.g.$ star 1 is being replaced by star 3
or star 1 being merged with star 3. But in the present work, we
always assume star 1 to be the neutron star, so to get binary
radio pulsar these latter two processes have to be rejected.

Interactions between binaries and single stars can enhance the
eccentricity of the binary orbit and this will be discussed in detail in the
subsequent sections. On the other hand, binary pulsars
emit gravitational waves which reduce both the size and eccentricity of the orbit
leading to mergers on a timescale of
$t_{gr}$. Following Peters \& Mathews (1963), $t_{gr}$ can be
calculated as follows :

\begin{equation}
t_{gr}= \left( \frac{1}{e}\frac{de}{dt}\right)^{-1}
\label{eq:tgr_e}
\end{equation}
where
\begin{equation}
\frac{de}{dt}=-\frac{304}{15}\frac{G^3 m_p m_c (m_p + m_c)}{c^5
a_p^4}g(e) \label{eq:dedt}
\end{equation}
and
\begin{equation}
g(e)=\left(1-e^2 \right)^{-5/2} e \left(1+\frac{121}{304}e^2
\right) \label{eq:ge}
\end{equation}
which is almost same as the expression of $t_{gr}$  calculated
from $da/dt$.

\begin{equation}
t_{gr}= \left( \frac{1}{a}\frac{da}{dt}\right)^{-1}
\label{eq:tgr_a}
\end{equation}
\begin{equation}
\frac{da}{dt}=-\frac{64}{5}\frac{G^3 m_p m_c (m_p + m_c)}{c^5
a_p^3}f(e) \label{eq:dadt}
\end{equation}
and
\begin{equation}
f(e)=\left(1-e^2 \right)^{-7/2} \left(1+\frac{73}{24}e^2 +
\frac{37}{96}e^4 \right) \label{eq:fe}
\end{equation}
In this work, we use eq. (\ref{eq:tgr_e}) to calculate $t_{gr}$.

%\subsection{Double neutron star binaries in globular clusters}
%\label{sec:dns}

%This

\subsection{Fly-by interactions}
\label{sec:fly_by}

\begin{figure}
{\includegraphics[width=0.5\textwidth,
height=0.3\textheight]{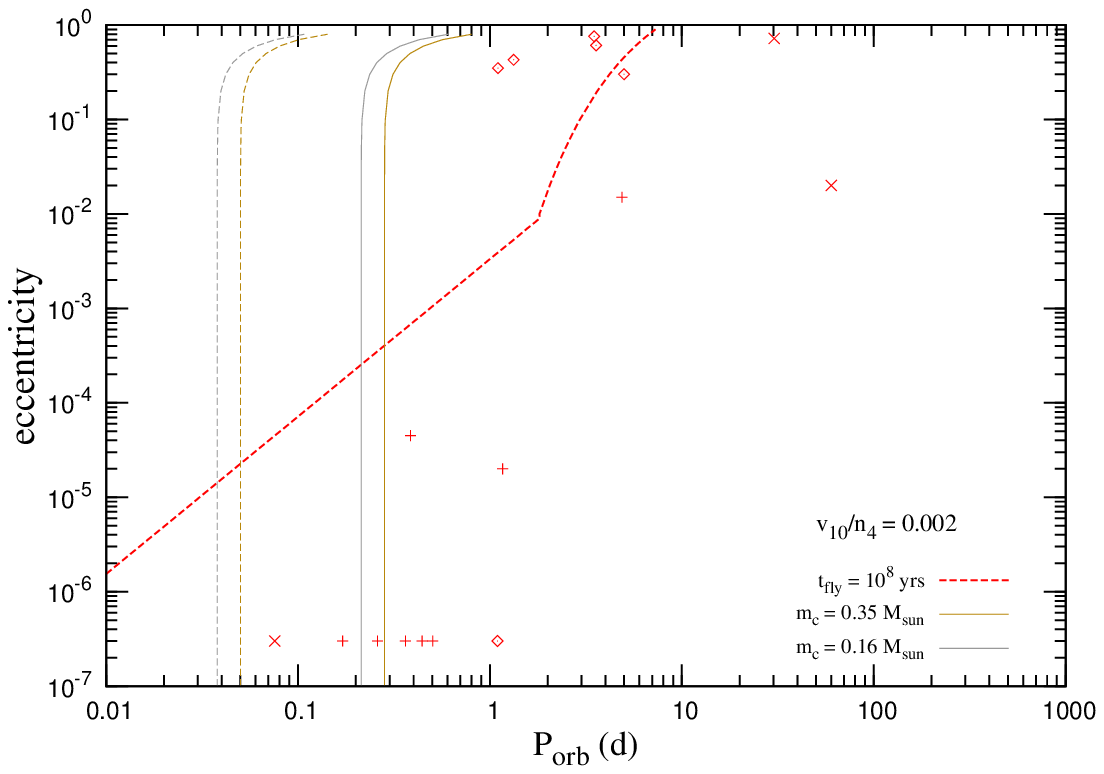}}
{\includegraphics[width=0.5\textwidth,
height=0.3\textheight]{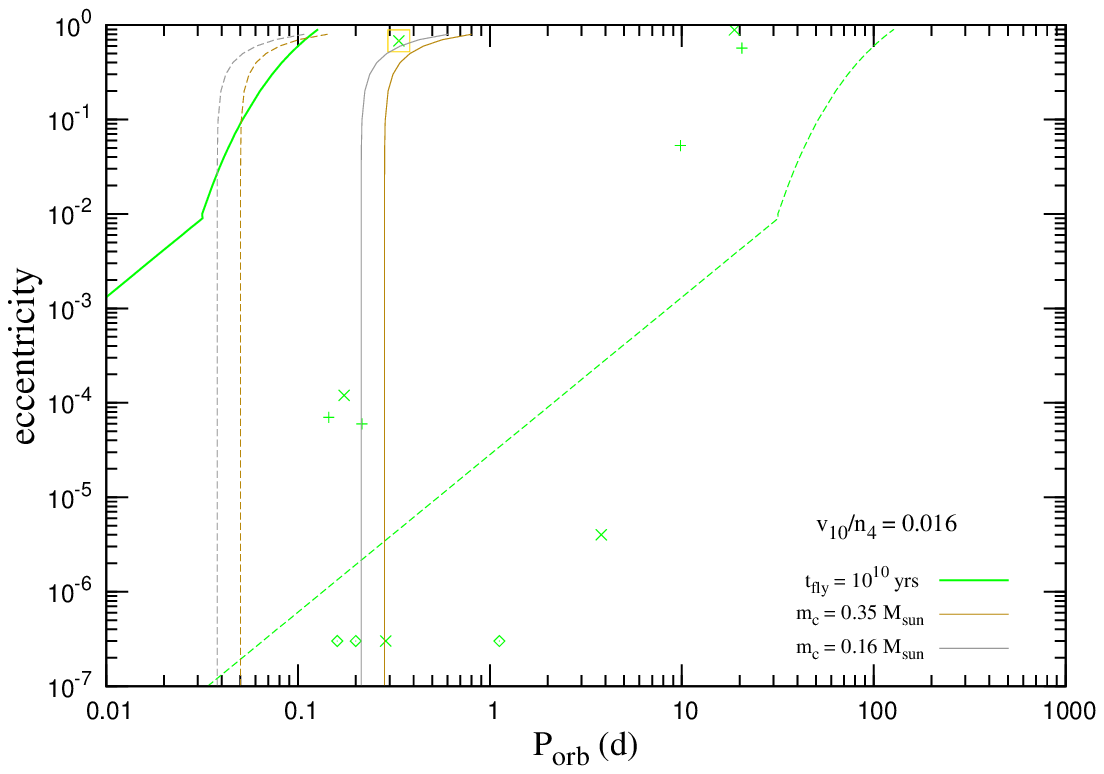}}
{\includegraphics[width=0.5\textwidth,
height=0.3\textheight]{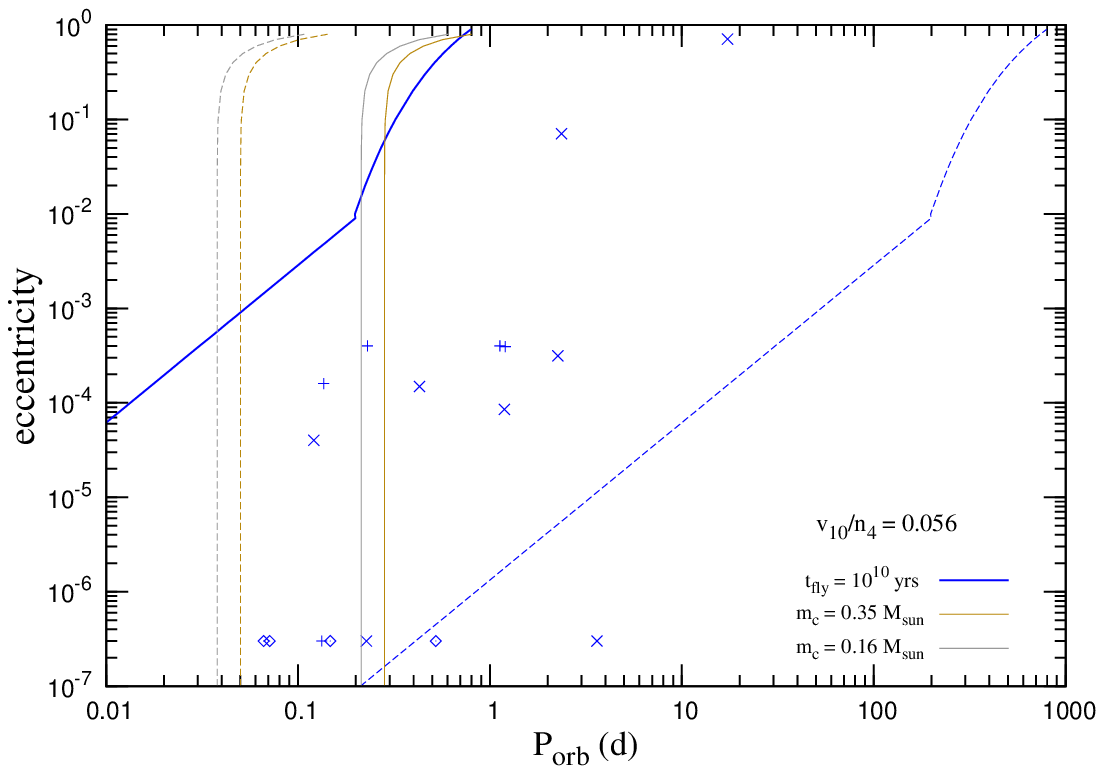}}
{\includegraphics[width=0.5\textwidth,
height=0.3\textheight]{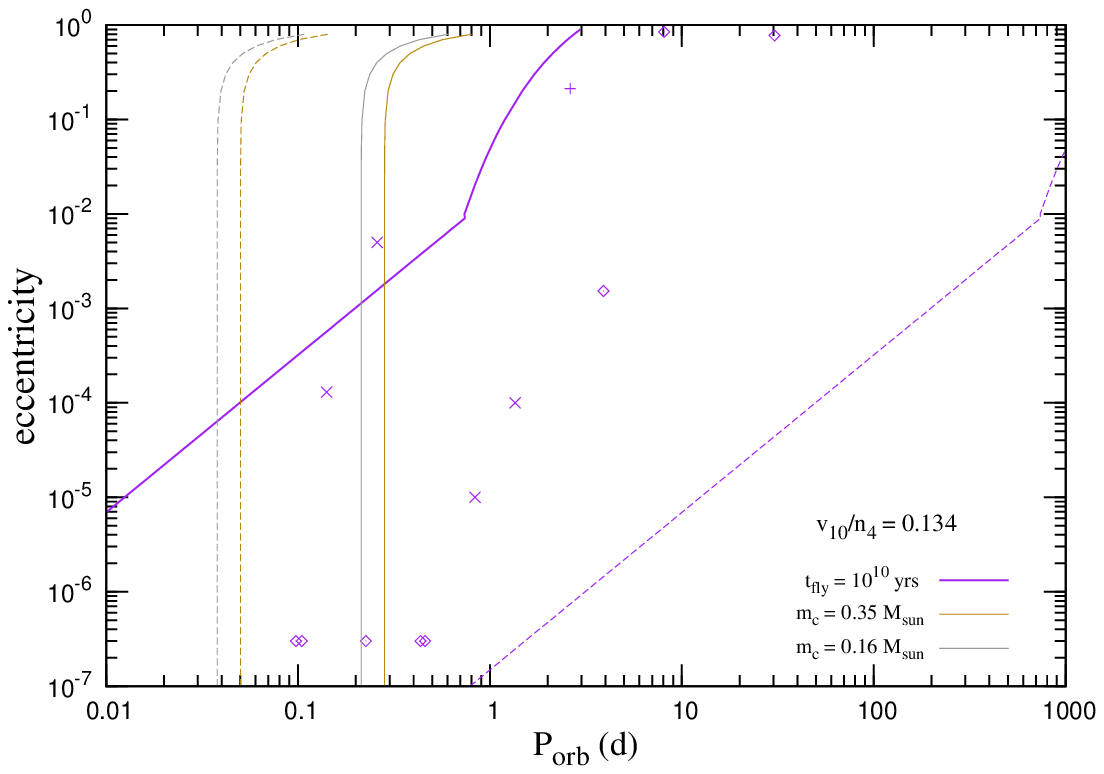}}
{\includegraphics[width=0.5\textwidth,
height=0.3\textheight]{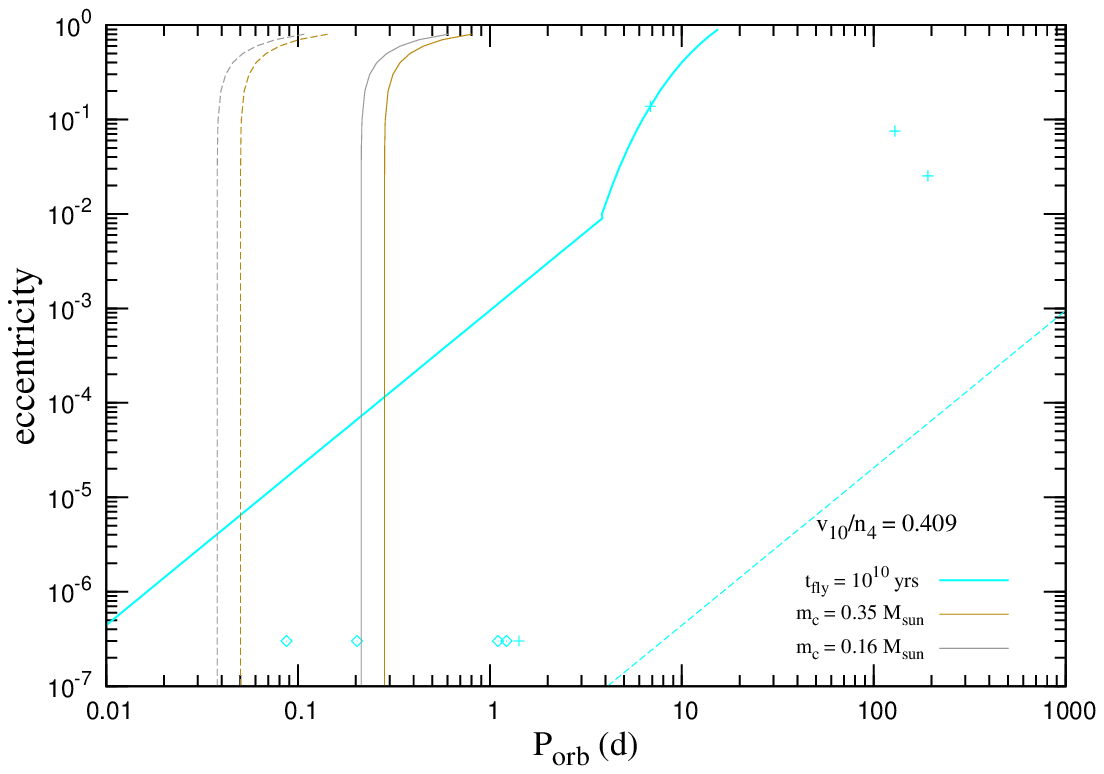}}
{\includegraphics[width=0.5\textwidth,
height=0.3\textheight]{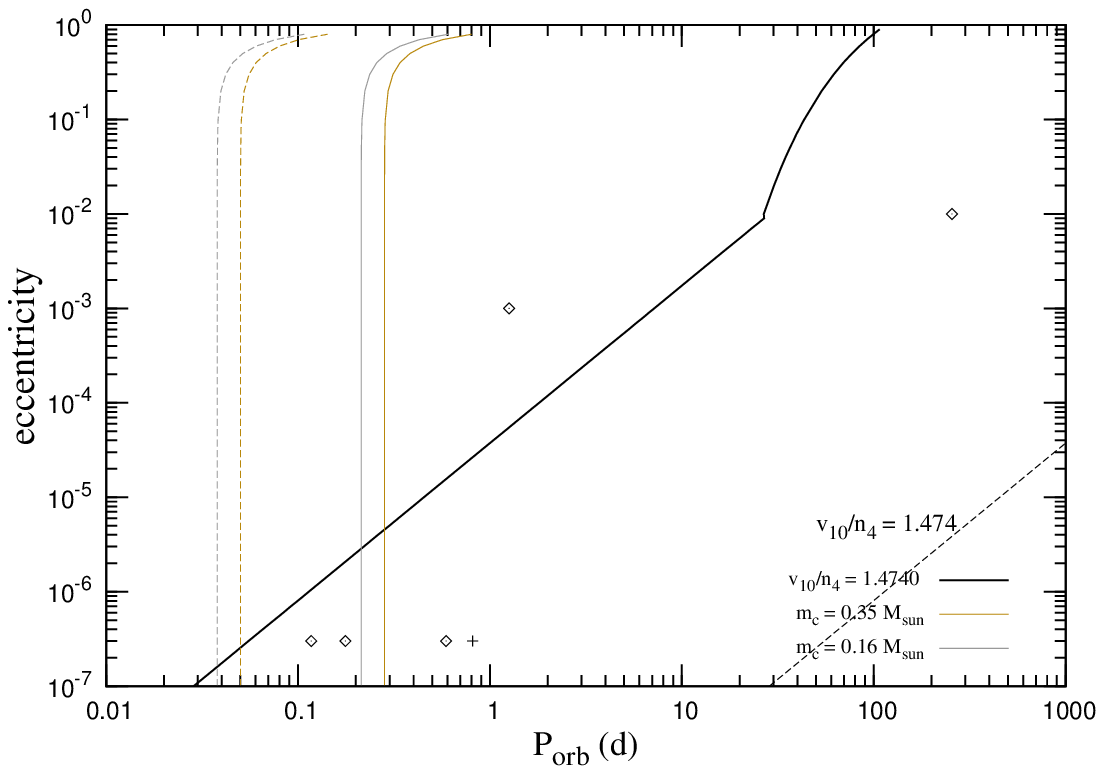}} \caption{Cluster binary pulsars
in the $e-P_{orb}$ plane with contours of $t_{fly}=10^{10}$ yrs
(solid lines) and $t_{fly}=10^{8}$ yrs (dashed lines) for each
group. Contours of $t_{gr}=10^{10}$ yrs (solid lines) and
$t_{gr}=10^{8}$ yrs (dashed lines) for binaries with $m_p~=~1.4~
M_{\odot}$ and $m_c~=~0.35~ M_{\odot}$ or $0.16~ M_{\odot}$ are
also shown. Pulsars with projected positions inside the cluster
core are marked with $+$, those outside the  cluster core are
marked with $\times$ and the pulsars with unknown position
offsets are marked with $\diamond$. Solid red line corresponding
to $t_{fly}=10^{10}$ years for $v_{10}/n_{4}= 0.0024$ is outside
the range plotted. Individual pulsars are marked with the same
colors as the $v_{10}/n_{4}$ values of the host GCs. }
\label{fig:psr_all_group_rasio_fly}
\end{figure}

Table \ref{tab:psr_parms} shows that most of the GC pulsars are
millisecond pulsars. Theoretically one expects that spun-up,
millisecond pulsars in binary systems formed from mass and angular
momentum transfer due to Roche lobe overflow and the resultant
tidal effects should appear in low eccentricity orbits $e \sim
10^{-6} - 10^{-3}$, \cite{phi92}. Since many highly eccentric
binary millisecond pulsars are found in globular clusters, this
indicates that stellar interactions are important for inducing
higher eccentricities. Rasio and Heggie \cite{ras95,heg96}
studied the change of orbital eccentricity of an initially
circular binary following a distant encounter with a third star
in a parabolic orbit. They used secular perturbation theory, i.e.
averaging over the orbital motion of the binary for sufficiently
large values of
 the pericenter distance $r_p$ where the encounter is quasi-adiabatic and used
non secular perturbation theory for smaller values
 of $r_p$ where the encounter is non-adiabatic. In the first case
$\delta e$ varies as a power law with $r_p/a_{in}$ and in the second case
$\delta e$ varies exponentially with  $r_p/a_{in}$. The power law dominates
for $e< 0.01$ and the exponential dominates for $e \gtrsim 0.01$. They used
the relation
$$b^2=r_p^2+2 G r_p (m_1+m_2+m_3)/v^2 \simeq 2G r_p (m_1+m_2+m_3)/v^2$$
to estimate the cross-sections
($\sigma=\pi b^2$) for eccentricity changes and then from $\sigma$, obtained
the time-scales for eccentricity changes as $t=1/(rate)=1/<n \sigma v >$
where $n$ is the number density. The timescales for these processes
are \cite{ras95}:

\begin{subequations}
\begin{eqnarray}
t_{fly}=4 \times 10^{11} n_4^{-1}v_{10}P_{orb}^{-2/3}e^{2/5}~ \rm {(e \lesssim 0.01) }\label{eq:tfly1}  \\
t_{fly}=2 \times 10^{11}
n_4^{-1}v_{10}P_{orb}^{-2/3}\left[-ln(e/4) \right]^{-2/3} ~\rm
{(e \gtrsim 0.01)}  \label{eq:tfly2}
\end{eqnarray}
\end{subequations}
where $n_4$ is the number density ($n$) of single stars in units of
$10^4~ \rm{pc^{-3} }$ and $v_{10}$ is the velocity dispersion ($v$) in units
of 10 km/sec in GCs, $P_{orb}$ is the orbital period in days giving $t_{fly}$
in years. The values of $v_{10}/n_{4}$ are different for different GCs
(table \ref{tb:gc_v_n}) and we grouped them into six different groups according
to their values of $v_{10}/n_{4}$ and calculated $t_{fly}$ for mean values
of $v_{10}/n_{4}$ for each group.
Cluster binary pulsars in the $e-P_{orb}$ plane with contours
of $t_{fly}=10^{10}$ yrs (solid lines) and $t_{fly}=10^{8}$ yrs (dashed lines)
for each group are shown in Fig \ref{fig:psr_all_group_rasio_fly}.
Contours of $t_{gr}=10^{10}$ yrs
(solid lines) and $t_{gr}=10^{8}$ yrs (dashed lines) for binaries with $m_p~=~1.4~ M_{\odot}$
and $m_c~=~0.35~ M_{\odot}$ or $0.16~ M_{\odot}$ are also shown.
Pulsars with projected positions inside
the cluster core are marked with $+$, those outside the  cluster core are marked with $\times$
and the pulsars with unknown position offsets are marked with $\diamond$.
Solid red line corresponding to
$t_{fly}=10^{10}$ years for $v_{10}/n_{4}= 0.0024$ is outside the range of axes used in our plots
(away from the upper left corner). Individual pulsars are marked with the same colors
as the $v_{10}/n_{4}$ values of the host GCs.
If a pulsar is located on the upper left half of the corresponding $t_{fly}=10^{10}$ years line $i.e.$ in the region where $t_{fly}>10^{10}$ years, then its eccentricity
cannot be due to fly-by interactions. There are two such binaries,
one is PSR B1718-19\footnote{Note however that its cluster association
is sometimes doubted, -- see \cite{cam05}, as it has a large offset from the GC center.}
(in NGC 6342, in group 4) and the other is PSR B 1639+36B (in M13, in group 6).
But the first one is a normal pulsar with $P_{s} = 1$ sec and as it is not spun up and may not
have been subject to a great deal of tidal forces, it can have escaped
circularization.
On the other hand, the second pulsar is a millisecond pulsar with
$P_{s} = 3.528$ ms, so it should have been in a circular orbit if it has not
had the time to go through any significant kind of stellar interactions.
Therefore it is of interest to investigate if it is a result of an exchange or
merger interaction. Moreover, some of the eccentric pulsars which seem to be
explainable by fly-by interactions lie outside of globular cluster cores (which
have less stellar density than in the core, and therefore less efficient
fly-by scattering). For a few
others positional offsets are not known. Even if a pulsar appears to be inside
the cluster core in the projected image, it can still be outside the cluster
core in three dimensional space. Globular cluster models show that values of
both  $v_{10}$ and $n_4$ fall outwards from cluster center and as the fall of
$n_4$ is much more rapid, the value of $v_{10}/n_4$ is higher outside the
cluster core. As we have already seen (different panels in Fig.
\ref{fig:psr_all_group_rasio_fly}) that an increase in the value of
$v_{10}/n_4$ shifts the contour of $t_{fly}~=~10^{10}$ years rightwards, this
will make the pulsar fall in the region where
$t_{fly}~>~10^{10}$ years excluding the possibility of fly-by interaction as the
cause of its eccentricity. In those cases also, we need to think about exchange
and/or merger interactions.
Another aspect of the pulsar population in these diagrams is that the long
orbital period pulsars are found predominantly in low density (and high
velocity dispersion) GCs, a feature that has been noticed before \cite{cam05}.
Camilo \& Rasio (2005) also point out that while
the LMXBs are predominantly found in very dense GCs, the radio pulsars tend to
be found more evenly distributed among low and high density clusters,
as also seen in the great deal of
variation of the $v_{10}/n_4$ parameter in the various plots of
Fig. \ref{fig:psr_all_group_rasio_fly}.

We have seen that there are many zero eccentricity pulsars which
lie in the region where $t_{fly}~<~10^{10}$ years. So it is of interest
to understand why they still appear in circular orbits despite the possibility
of fly-by interactions.
It is possible however that, firstly the binaries may be very young so that
they have not had enough time to interact with single stars. Secondly, they may
in fact lie outside cluster cores with a higher $v_{10}/n_4$ even though in
projection many of them may appear to be
inside the cores so that the timescale for
fly-by is actually $t_{fly}~>~10^{10}$ years.

Positions of cluster binaries with respect to the cluster center
in the projected plane is given in Table \ref{tab:psr_parms}.
There are a number of pulsars for which position
offsets are not known. Also the binaries which appear to be
inside the cluster core in the projected plane can actually lie
outside the core.
The observed values of spin period derivatives ($\dot{P_{s}}$) of
the globular cluster pulsars can give some indication of the pulsar position
and its environment since
negative or positive $\dot{P_{s}}$ indicate positions in the back half or in
the front half of the cluster respectively \citep{phi92}; however, the
actual position cannot be determined with the knowledge of $\dot{P_{s}}$ only.
DM measure could also be a tool to measure pulsar positions with respect to
cluster center if DM variations were dominated by
intra-cluster gas \citep{ransom06}. However, it has been shown that
for Terzan 5, DM is mainly due to ISM (Ransom2006), so DM values alone
can not help to determine pulsar positions with respect to the center
for this cluster.

\subsection{Exchange and merger interactions}
\label{sec:starlab}

Since exchange and merger interactions have not been amenable to
analytic treatment, we performed numerical scattering experiments using
STARLAB codes\footnote{www.ids.ias.edu/$\sim$starlab/}.
We used the STARLAB task ``sigma3" which gives the scaled cross section $X$ for
different types of interactions, $e.g.$
fly-by, exchange, merger and ionization (both resonant and non-resonant).
The inputs given are : velocity ($v$) of the incoming star
(with mass and radius $m_3, R_3$) and the masses, the radii and the semi major axis
of the initial binary ($m_1, R_1;~ m_2, R_2;~a_{in}$). $X$ is defined as :
\begin{equation}
X~=~\frac{\sigma}{\pi~a_{in}^2}\left(\frac{v}{v_c} \right)^2
\label{eq:X}
\end{equation}
where $\sigma$ is the cross section and $v_c$ is the critical velocity as
defined earlier (eq. \ref{eq:vc}).

We select the range of $a_{in}$ as $0.001- 8.0~AU$ so that for all sets of
stellar parameters, the initial orbital periods are well within the range of
observed values of orbital periods $0.01 - 1000$ days.
But depending upon the stellar parameters, one or more values of $a_{in}$ are
automatically rejected by STARLAB whenever $a_{in}$ is too low
(giving ``contact binaries"). We divide the range of $a_{in}$ in 35 bins
($N_{bin}$) equal in logarithmic scales to invoke ``sigma3". For each value
of $a_{in}$, we take the maximum trial density ($n_{trial}$) as 5000 which
is uniformly distributed in the impact parameter ($\rho$) over the range
$ 0 \leq \rho \leq \rho_0$ where $\rho_0$ simply corresponds to a periastron
separation of $2a$. The impact parameter range is then systematically expanded
to cover successive annuli of outer radii $\rho_i = 2^{i/2} \rho_0$ with
$n_{trial}$ trials each until no interesting interaction take place in the
outermost zone \citep{mcm96}. Thus the total number of trials become
significantly large. As an example, for the parameters $m_1~= 1.4~M_{\odot},
R_1~=10 ~{\rm km };~ m_2~= 0.16~M_{\odot}, R_2~= 0.16~R_{\odot};~ m_3~=0.33~M_{\odot},
R_3~=~R_{\odot};~a_{in}~=~0.2 {\rm AU}; ~v~= 11.76~{\rm km /sec}$
we had a sample total of 15570 scatterings.
Out of these 10960 were fly-bys, 3722 exchanges, 982 two mergers and
6 three-mergers.
The CPU time needed to perform this particular set of
interaction was 1171.23 secs\footnote{The simulations were performed on a
HP Proliant BL465C computer with Dual-Core 2.6 GHz AMD Opteron
Processor 2218 with 16 GB RAM.}.
Though widely different values of $v$ were chosen (13.15 km/sec, 11.76 km/sec,
7.79 km/sec and 3.29 km/sec) covering the entire range of $v$ in 26 GCs, we
concentrated mainly on the runs for $v=11.76$ km/sec corresponding to the
velocity dispersion relevant to Terzan 5, the host of the largest number of
known binary radio pulsars. Different parameters for STARLAB runs performed
for this value of $v$ are given in Table \ref{tb:starlab_runs}.

\begin{figure}
{\includegraphics[width=0.5\textwidth,
height=0.3\textheight]{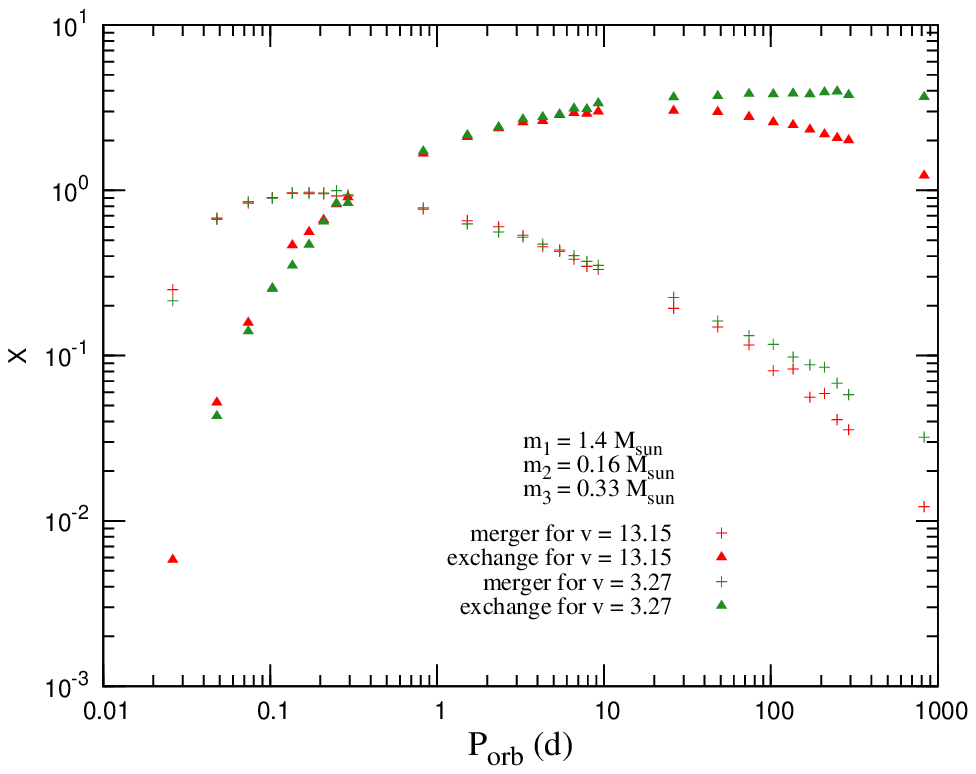}}
{\includegraphics[width=0.5\textwidth,
height=0.3\textheight]{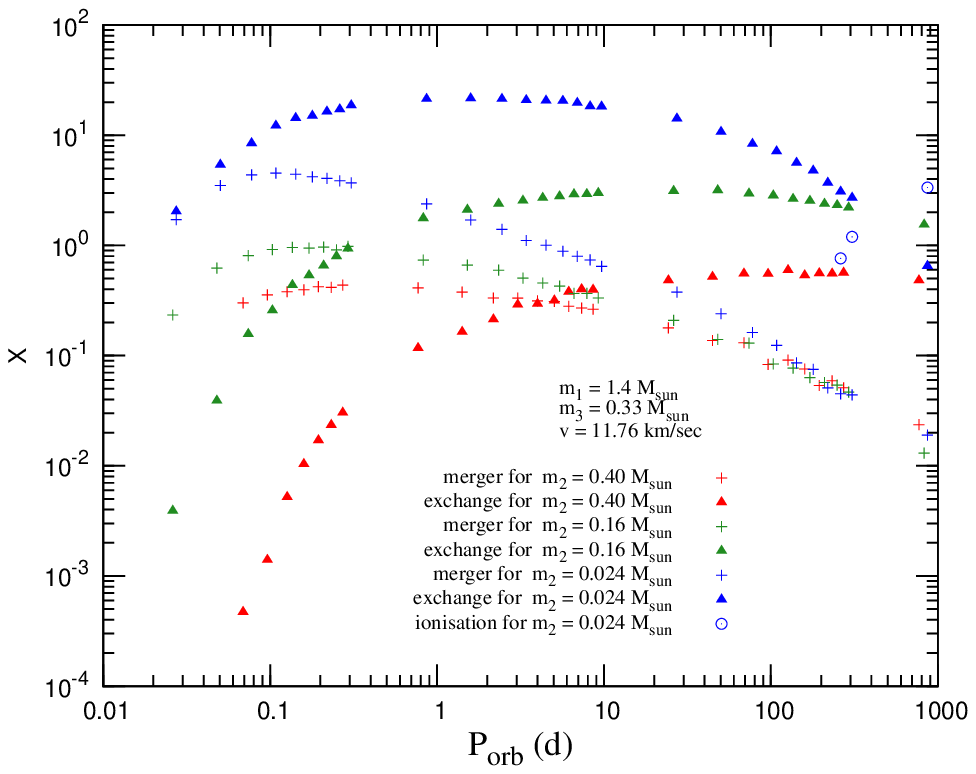}}
{\includegraphics[width=0.5\textwidth,
height=0.3\textheight]{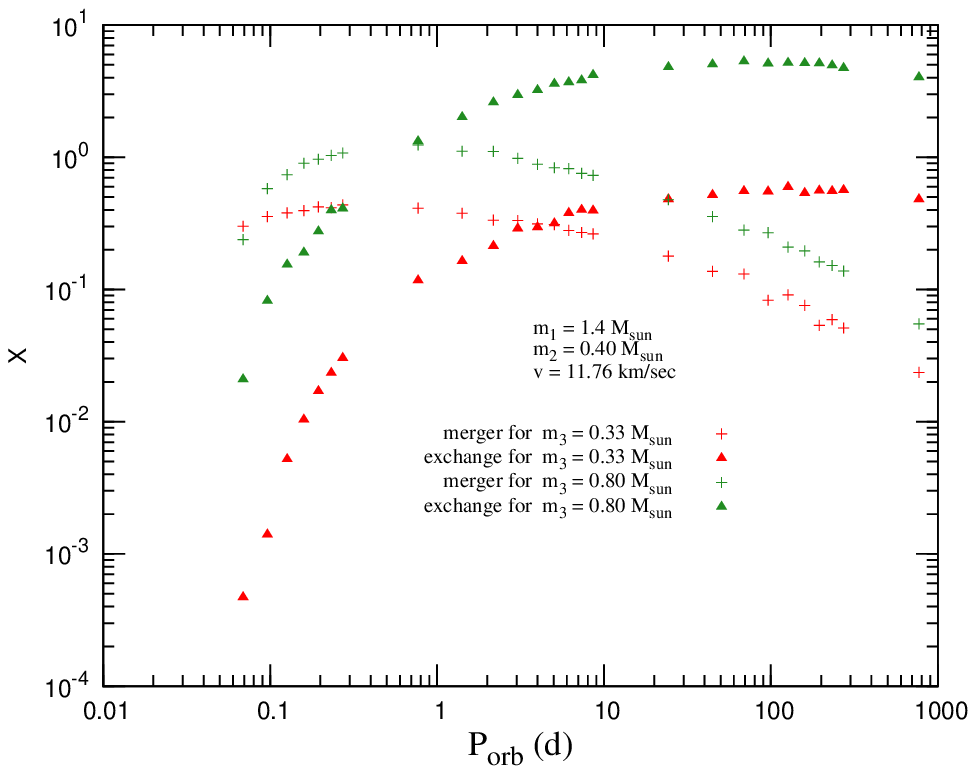}}
{\includegraphics[width=0.5\textwidth,
height=0.3\textheight]{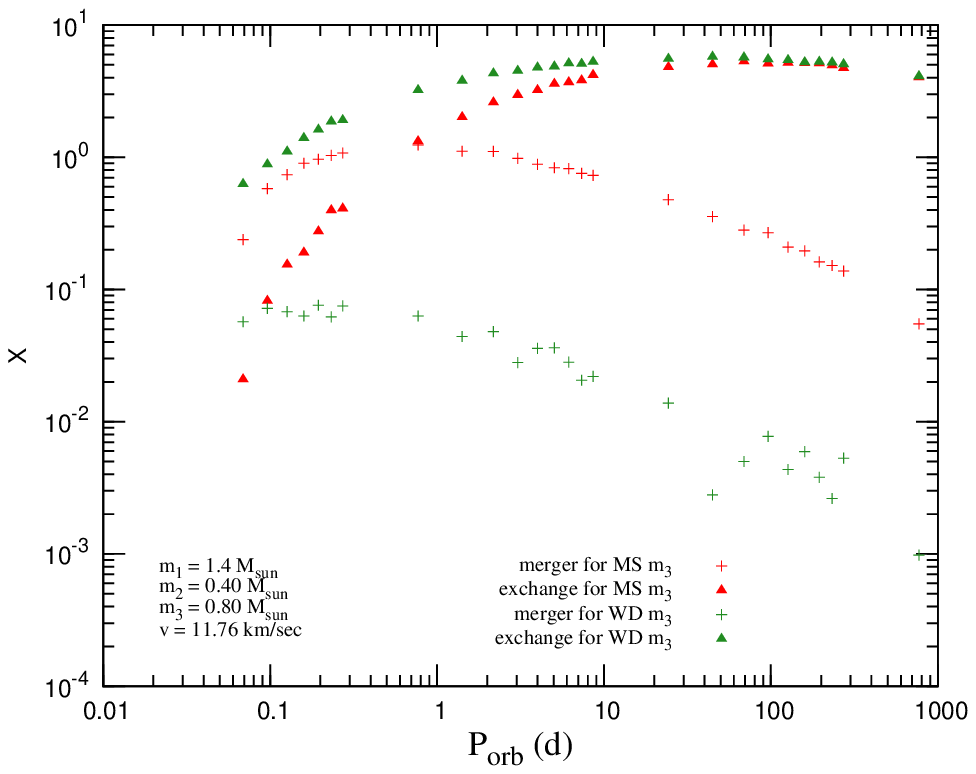}} \caption{Variation of $X$ for
exchange ($+$), merger ($\blacktriangle$) and ionization
($\circ$, whenever significant) with $P_{orb,~in}$. (i) The first
figure is for $m_1=1.4~M_{\odot},~m_2=0.16~M_{\odot},
~m_3=0.33~M_{\odot}$ and two values of $v$ $e.g$ 13.16 km/sec
(red) and 3.27 km/sec (green).  (ii) The second figure is for
$m_1=1.4~M_{\odot},~m_3=0.33~M_{\odot},$ $v=11.76$ km/sec and
different values of $m_2$ $e.g$ $0.40~M_{odor}$ (red),
$0.16~M_{\odot}$ (green) and $0.024~M_{\odot}$ (blue). (iii) The
third figure is for $m_1=1.4~M_{\odot},~m_2=0.16~M_{\odot},$
$v=11.76$ km/sec and different values of $m_3$ $e.g.$
$0.33~M_{\odot}$ (red), $0.80~M_{\odot}$ (green).  (iv) The fourth
figure is for $m_1=1.4~M_{\odot},~m_2=0.16~M_{\odot},
~m_3=0.80~M_{\odot}$, $v=11.76$ km/sec and  the $3^{rd}$ star is
either a MS (red) or a WD (green).} \label{fig:X_porb}
\end{figure}

For each set of $a_{in}$, initial orbital period $P_{orb,~in}$ is
calculated using Kepler's law as stellar parameters are known.
STARLAB outputs contain X corresponding to each $a_{in}$ $i.e.$
corresponding to each $P_{orb,~in}$ for all types of
interactions. In Fig. \ref{fig:X_porb} we plot the variation of
$X$ for exchange ($+$), merger ($\blacktriangle$) and ionization
($\circ$, whenever significant) with $P_{orb,~in}$ and check the
dependencies on different parameters - (i) In the first figure,
we set $m_1~=~1.4~M_{\odot},~m_2~=~0.16~M_{\odot}$,
$m_3~=~0.33~M_{\odot}$, and take two values of $v$ $e.g.$ 13.16
km/sec (red) and 3.27 km/sec (green) which are the highest and
the lowest values (see Table \ref{tb:gc_v_n}). A change in the
value of $v$ does not change the value of $X$ much when the other
parameters are kept fixed. (ii) In the second figure, we set
$m_1~=~1.4~M_{\odot},~m_3~=~0.33~M_{\odot},$ $v~=11.76$ km/sec,
and take different values of $m_2$ $e.g.$ $0.40~M_{\odot}$ (red),
$0.16~M_{\odot}$ (green) and $0.024~M_{\odot}$ (blue). Both
$X_{exchange}$ and $X_{merger}$ increases with decrease of $m_2$.
Ionization starts for $m_2=~0.024~M_{\odot}$ at
$P_{orb,~in}\approxeq 200$ days.  (iii) In the third figure, we
set $m_1~=~1.4~M_{\odot},~m_2~=~0.16~M_{\odot},$ $v~=11.76$
km/sec, and take different values of $m_3$ $e.g.$ $0.33~M_{\odot}$
(red), $0.80~M_{\odot}$ (green). Both $X_{exchange}$ and
$X_{merger}$ increases with increase of $m_3$. (iv) In the fourth
figure, we set $m_1~=~1.4~M_{\odot},~m_2~=~0.16~M_{\odot},
~m_3~=~0.80~M_{\odot}$, $v~=~11.76$ km/sec and take the $3^{rd}$
star to be either a MS (red) or a WD (green). For WD,
$X_{exchange}$ is higher but $X_{merger}$ is lower.

\begin{figure}
{\includegraphics[width=0.5\textwidth,
height=0.35\textheight]{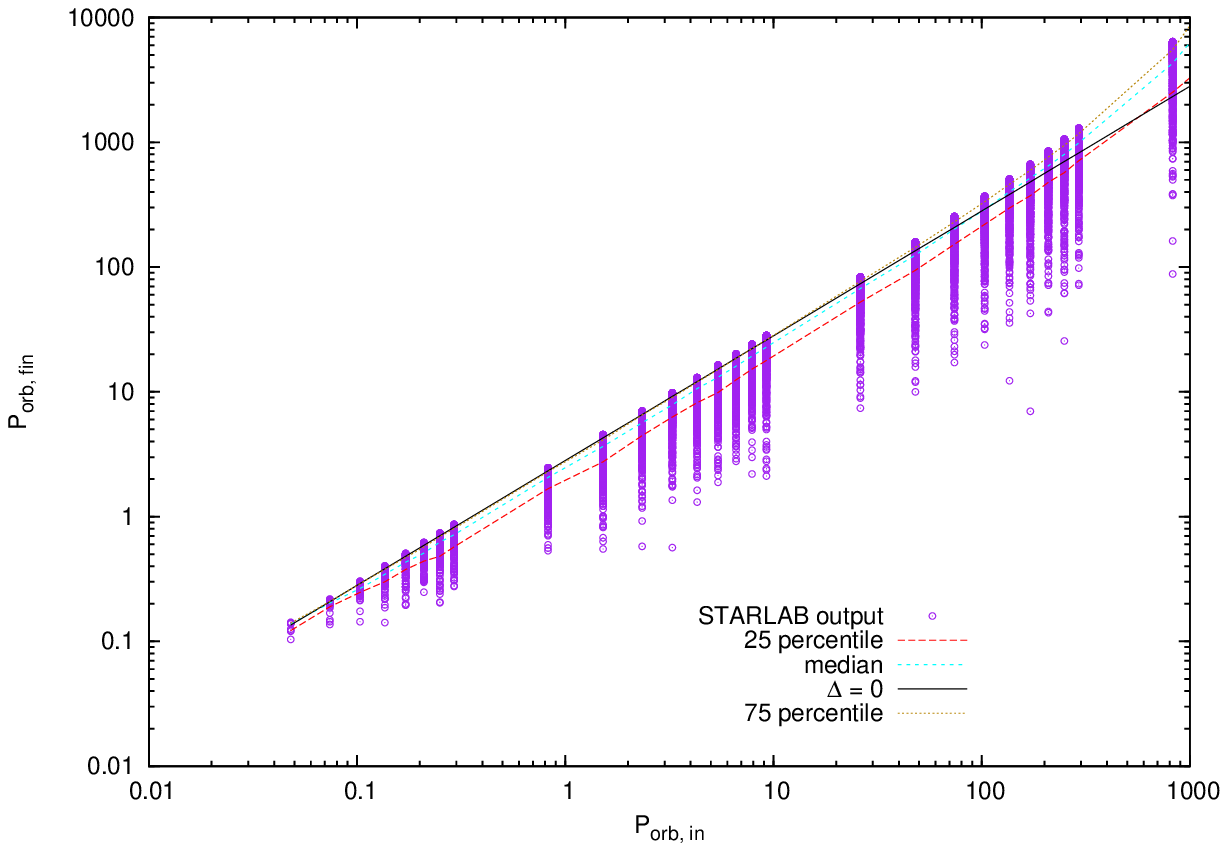}}
{\includegraphics[width=0.5\textwidth,
height=0.35\textheight]{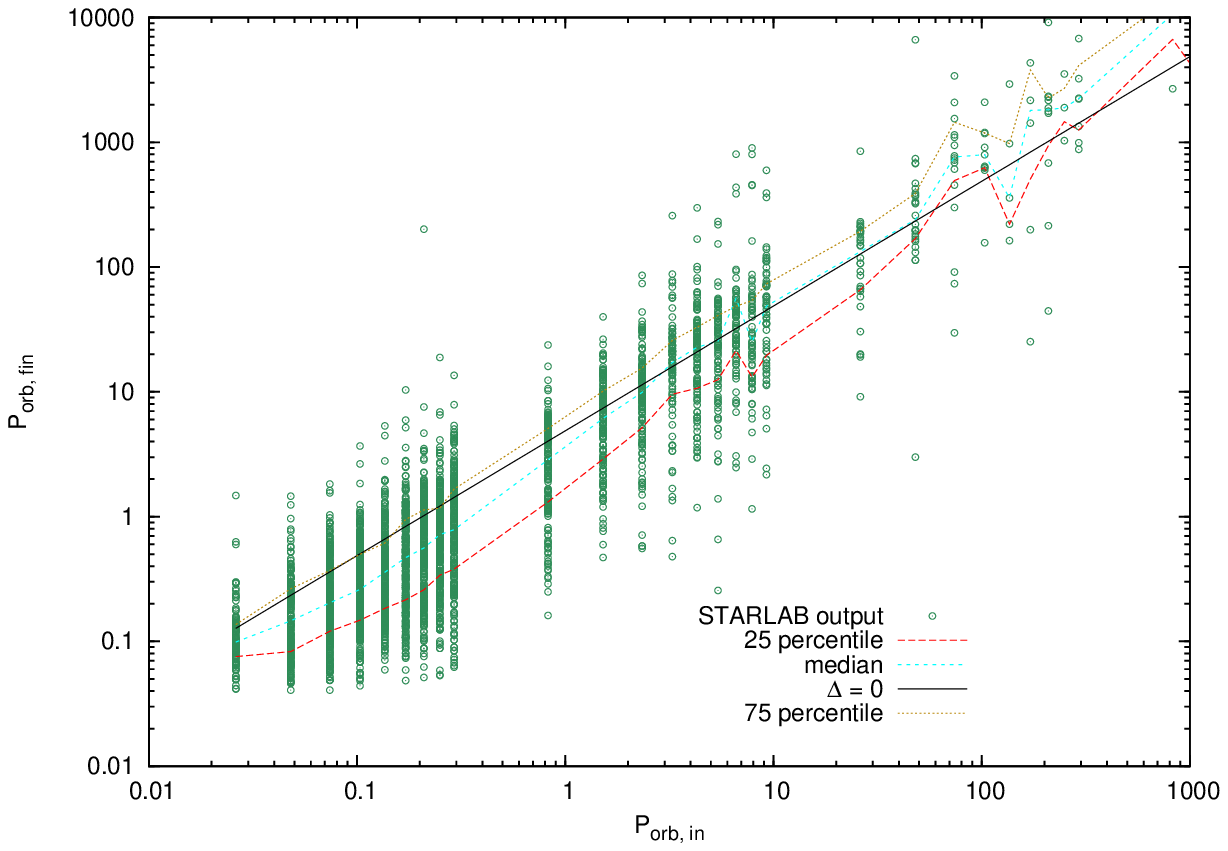}} \caption{$P_{orb,~in}$ Vs
$P_{orb,~fin}$ plots. Right hand side (purple points) diagram is
for exchanges and the left hand side diagram (green points) is
for mergers. The stellar parameters are as follows :
$m_1=1.4~M_{\odot}$ (NS), $m_3=0.33~M_{\odot}$ (MS),
$m_2=~0.16~M_{\odot}$ (MS) and $v=11.76$ km/sec.}
\label{fig:pin_pfin_stat}
\end{figure}

From $X$, one can calculate the cross-section ($\sigma$  see eq. \ref{eq:X}) and the
interaction time scale $t$ as $t = 1 / n \sigma v$ where $n$
is the number density of single stars. STARLAB also gives the properties of final states, $i.e.$ eccentricities and semi major axes of
the final binaries
for each set of inputs ($m_1, R_1;~ m_2, R_2;~ m_3, R_3;~a_{in}$). So for each
value of $P_{orb,~in}$, we get a number of values of $P_{orb,~fin}$ and $e_{fin}$.
For each value of $P_{orb,~in}$, we calculated $25$ percentile,
median and  $75$ percentile values of $P_{orb,~fin}$ as well as the values from
analytic expressions given in eq. \ref{eq:afin_ex} and \ref{eq:afin_merg}
putting $\Delta=0$. As all these values are very close to each other,
we use the $P_{orb,~in}-P_{orb,~fin}$ relation corresponding to $\Delta~=~0$
throughout. As an example, in Fig. \ref{fig:pin_pfin_stat}, we have plotted
$P_{orb,~in}$ Vs $P_{orb,~fin}$ as reported by STARLAB (scatter plots)
for both merger and exchange with
$m_1~=~1.4~M_{\odot},~m_2~=~0.16~M_{\odot}, ~m_3~=~0.33~M_{\odot}$, $v~=~11.76$ km/sec.
The line for 25 percentile,
median, 75 percentile along with the analytic relation of
$P_{orb,~in}-P_{orb,~fin}$ with $\Delta~=~0$ (eq. \ref{eq:afin_gen}) are shown
in the same plot.

\begin{figure}
{\includegraphics[width=0.5\textwidth,
height=0.3\textheight]{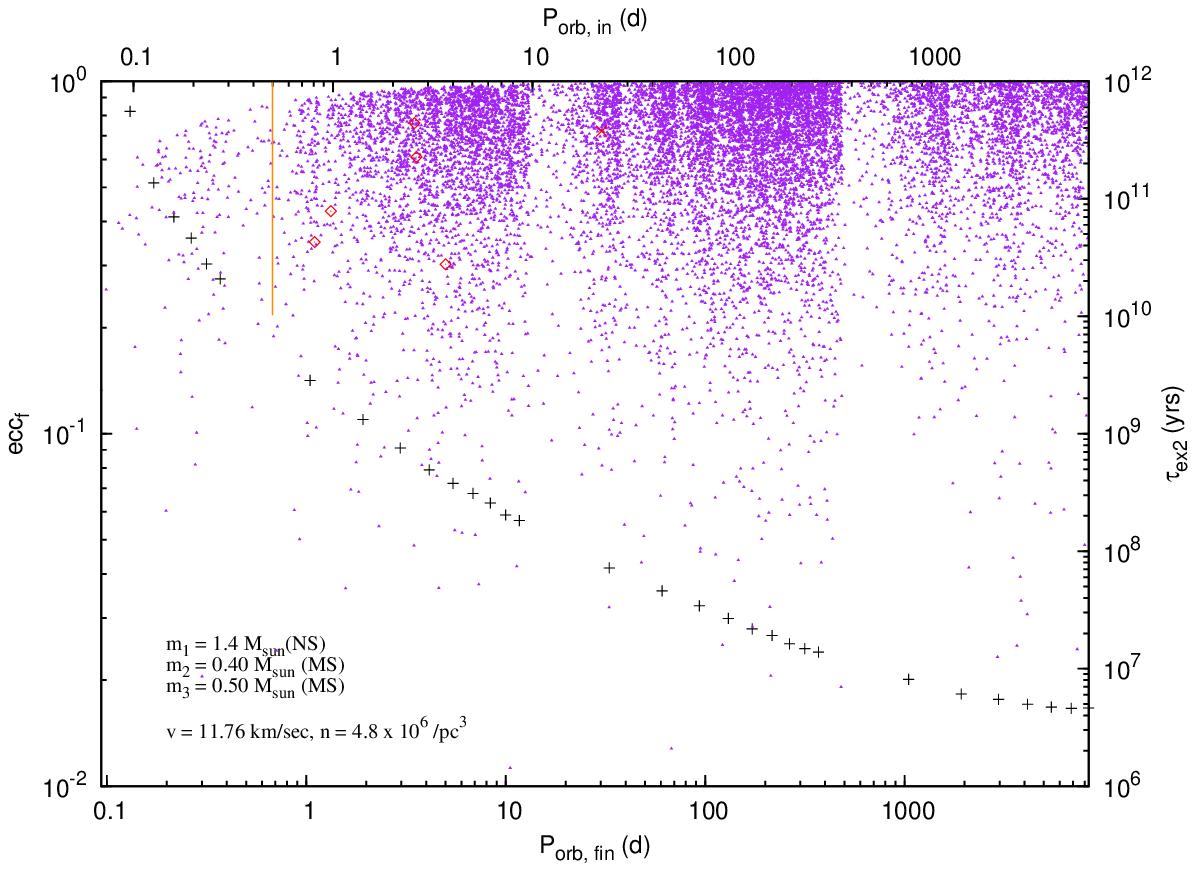}}
{\includegraphics[width=0.5\textwidth,
height=0.3\textheight]{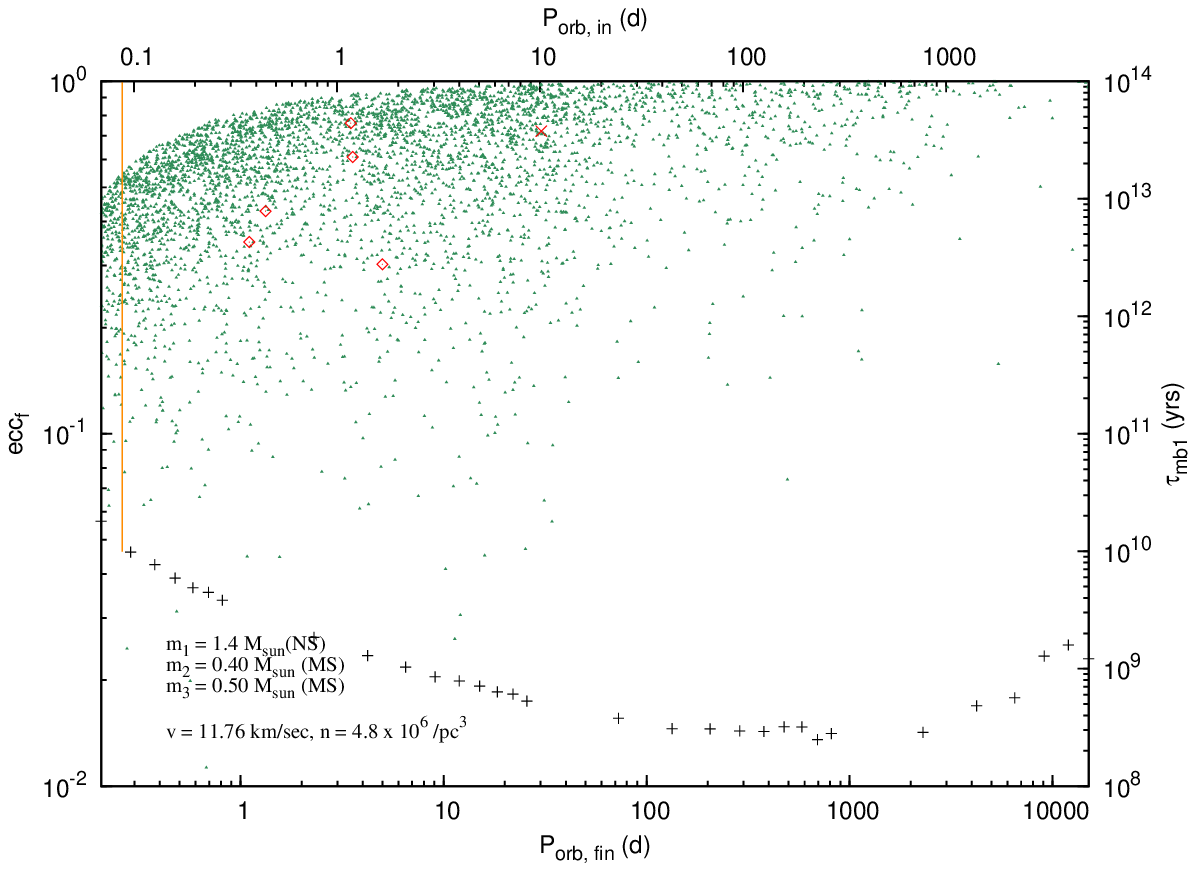}}
{\includegraphics[width=0.5\textwidth,
height=0.3\textheight]{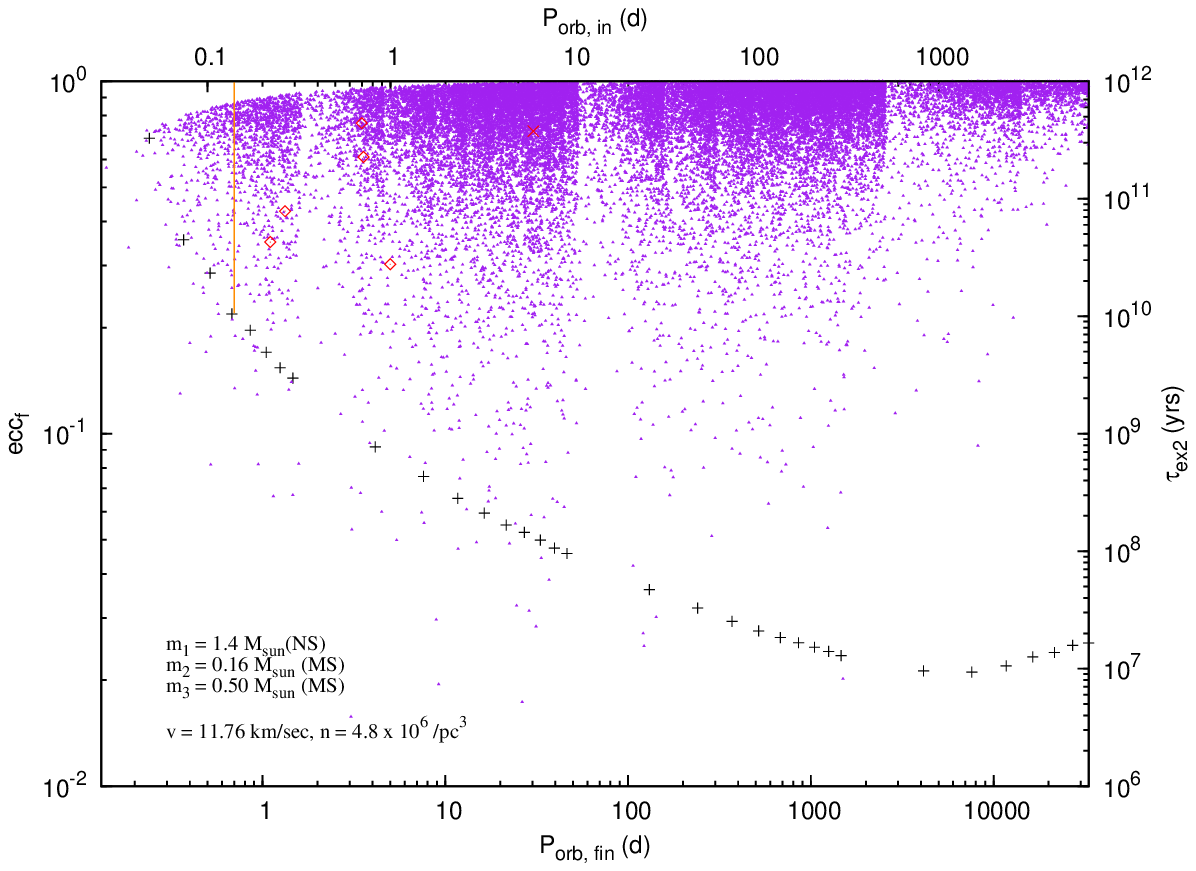}}
{\includegraphics[width=0.5\textwidth,
height=0.3\textheight]{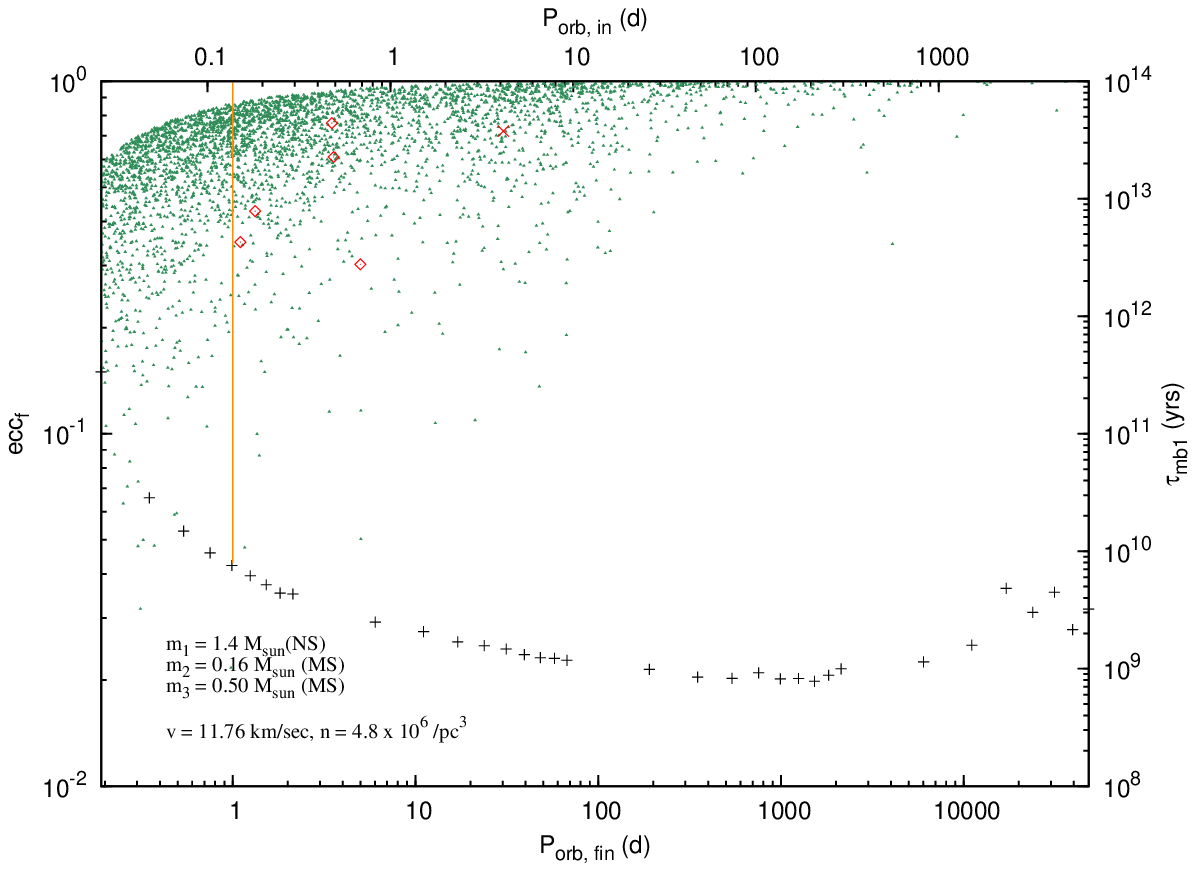}}
{\includegraphics[width=0.5\textwidth,
height=0.3\textheight]{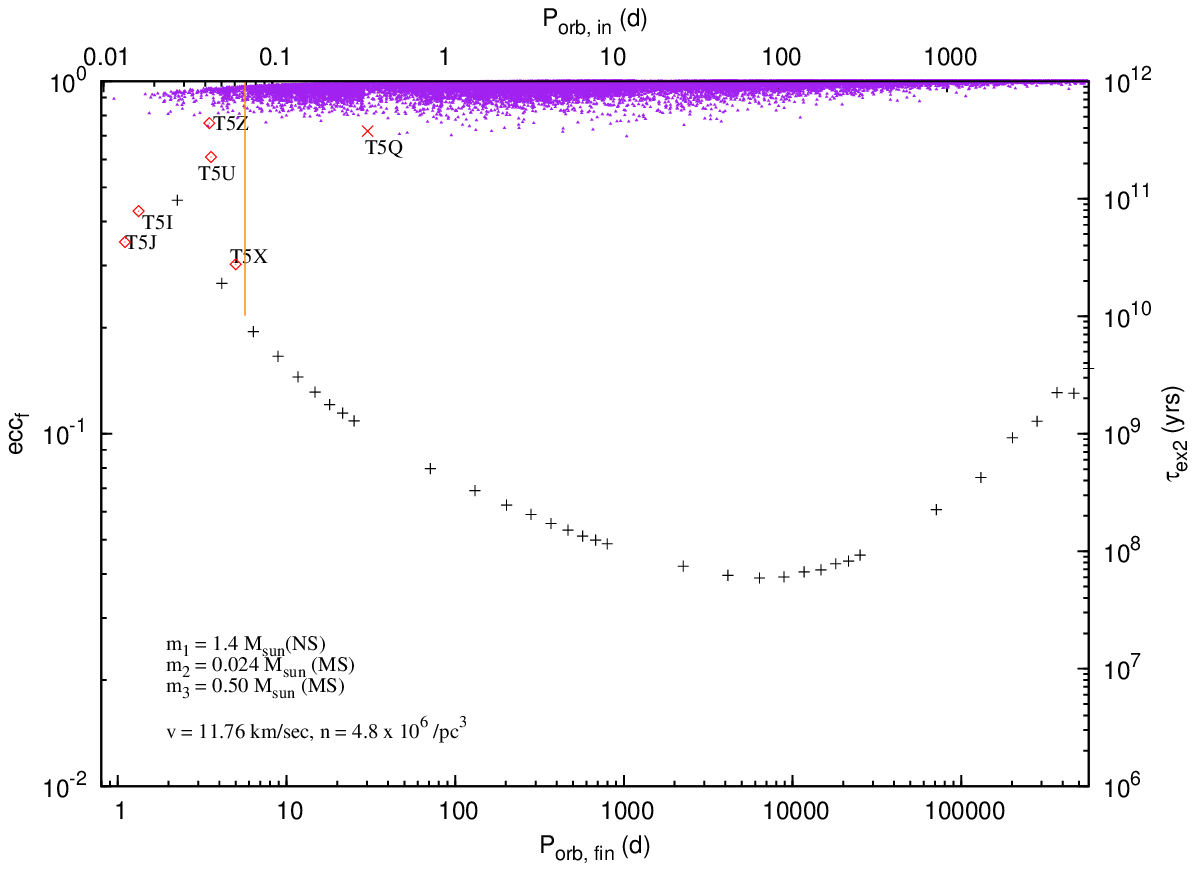}}
{\includegraphics[width=0.5\textwidth,
height=0.3\textheight]{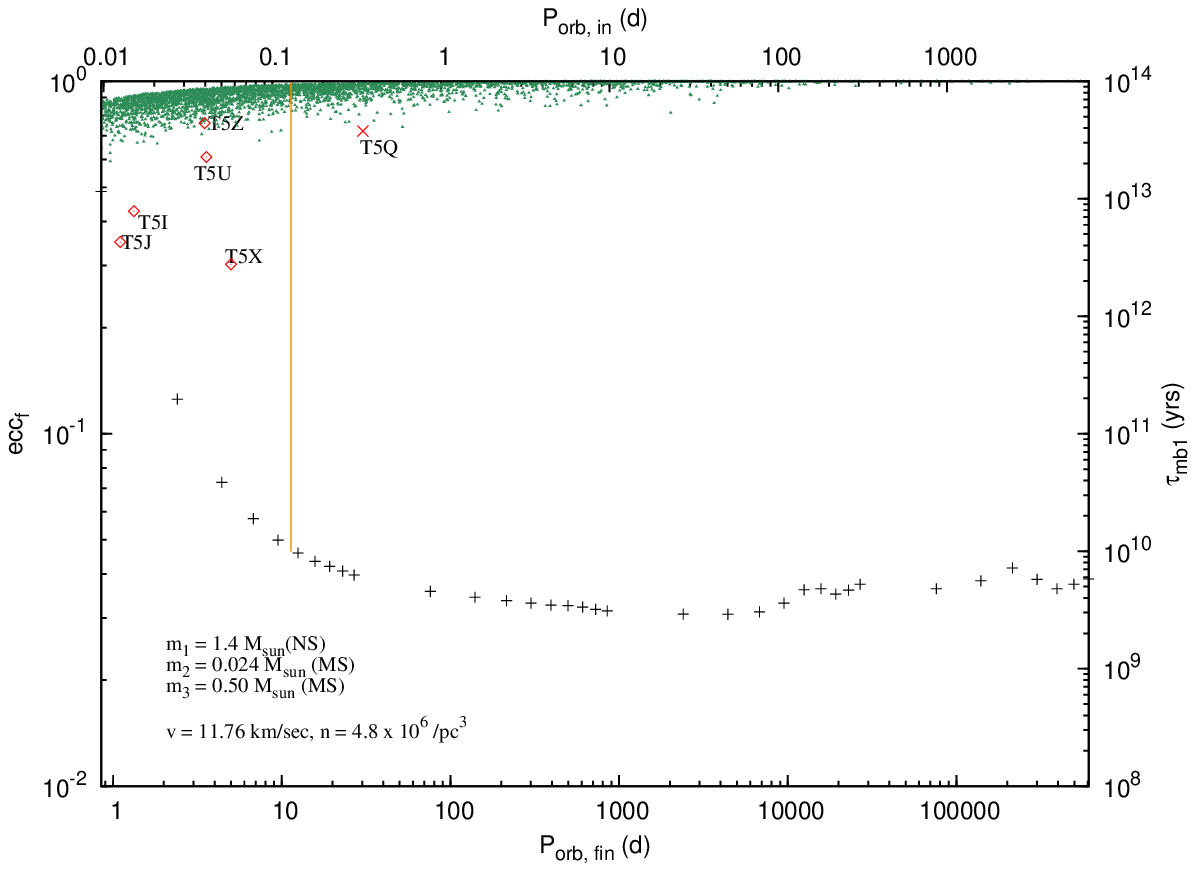}} \caption{\footnotesize{Time
scales (denoted by `+') and final eccentricity distributions
(scatter-plot of points) with initial and final orbital periods
($\Delta~=~0$ in eq. \ref{eq:afin_ex}) for exchange (red points:
`` exchange2") and merger (green points:  `` merger$\_~$ b1")
interactions with different stellar parameters. We plot $P_{orb,
in}$  along the top x-axis and $P_{orb, fin}$  along the bottom
x-axis. The left y axis gives the final eccentricities while the
right y axis gives the time scales of interactions. The vertical
orange lines form the boundaries of the allowed orbital period
regions where interaction time scales are less than $10^{10}$
yrs. The stellar parameters are as follows : $m_1=1.4~M_{\odot}$,
$m_3=0.50~M_{\odot}$ and $m_2$ varies from $0.024~M_{\odot}$ to
$0.16~M_{\odot}$ and then to $0.40~M_{\odot}$. }}
\label{fig:terzan_exch_merg_m3_h}
\end{figure}

In Fig. \ref{fig:terzan_exch_merg_m3_h},
we plot $P_{orb, in}$ of the initial binary (comprising of stars $m_1$ and $m_2$)
along the top x-axis and $P_{orb, fin}$  along the bottom x-axis.
Purple points in the left panels are for exchange interactions while the green points
in the right panel are for the merger interactions. The left y axis gives the final
eccentricities while the right y axis gives the time scales of the
interactions. Interaction time scales are plotted with black $+$s.
The vertical orange lines form the boundaries of the allowed orbital period regions
where interaction time scales are less than $10^{10}$ years. It is clear from the
scatter plots (Fig. \ref{fig:terzan_exch_merg_m3_h}) that the final binaries will most probably have $e>0.1$ if they undergo either exchange or merger events.
The observed pulsars with $e>0.1$ in Terzan 5 can be found in
Table \ref{tab:psr_parms} where the companion masses ($m_{c}$) are for inclination
angle $i~=~60^{\circ}$ and are also shown in the scatter plots
(red in color, named in few plots). In Fig \ref{fig:terzan_exch_merg_m3_h}, $m_1=1.4~M_{\odot}$, $m_3=0.50~ M_{\odot}$ and
we vary $m_2$ from $0.024~ M_{\odot}$ to $0.16~ M_{\odot}$ and $0.40~ M_{\odot}$. In Table \ref{tb:starlab_runs}, we summarize
the stellar parameters for the different sets of STARLAB runs.

If an observed eccentric binary pulsar lies in the region where time scale for
a particular interaction is greater than $10^{10}$ years, then
that interaction can not be responsible for its eccentricity.
On the other hand, an exchange interaction to be the origin of the eccentricity of
a particular pulsar, we should have $m_c \approxeq m_3$ and for merger $m_c \approxeq m_2+m_3$. Considering these two facts, we can surmise that
PSR I may have resulted from exchange interaction
$m_2$ is either $0.16~M_{\odot}$ or $0.40~M_{\odot}$, PSR Q may have resulted from merger interaction when $m_2$ is $0.16~M_{\odot}$, PSR U can result from exchange interaction when $m_2$ is either $0.16~M_{\odot}$ or $0.40~M_{\odot}$ or
from merger interaction when $m_2$ is $0.16~M_{\odot}$, PSR Z can result from merger interaction when $m_2$ is  either $0.16~M_{\odot}$ $0.40~M_{\odot}$ \footnote{Comparing Fig \ref{fig:terzan_exch_merg_m3_h} of the present manuscript with the Fig 2 of Bagchi \& Ray (2009), it is clear that with a change of the value of $m_3$ from 0.33 $M_{\odot}$ to 0.50 $M_{\odot}$, the outcomes of the simulations i.e. the appearance of the scatter plots does not change much in comparison to its change with the change in the value of $m_2$ (for fixed $m_3$). For the sake of completeness, we wish to remind the readers that if $m_3$ is $0.33~M_{\odot}$, then PSR I may have resulted from exchange interaction when $m_2$ is either $0.16~M_{\odot}$ or $0.40~M_{\odot}$, or merger interaction when $m_2$ is $0.16~M_{\odot}$, PSR J could have resulted from exchange interaction when $m_2$ is either $0.16~M_{\odot}$ or $0.40~M_{\odot}$, PSR Q may have resulted from merger when $m_2$ is $0.40~M_{\odot}$, PSR U can result from merger interaction when $m_2$ is $0.40~M_{\odot}$, PSR X can result from exchange interaction when $m_2$ is either $0.16~M_{\odot}$ or $0.40~M_{\odot}$, PSR Z can result from merger interaction when $m_2$ is  $0.40~M_{\odot}$.}.
These conditions on stellar masses are mainly indicative and need not be satisfied very accurately
because: 1) a slightly different
 choice of pulsar masses can give the same or similar
output of the simulation and
2) the companion masses are not known exactly in most cases; the masses
are obtained
from the mass function in terms of $m_c sin\;i$ and with the assumption that
the orbital inclination angle
is $60^{\circ}$ (Table \ref{tab:psr_parms}).

\section{Ionization}
\label{sec:ion}

\begin{figure}[h!]
\epsscale{1.0} \plotone{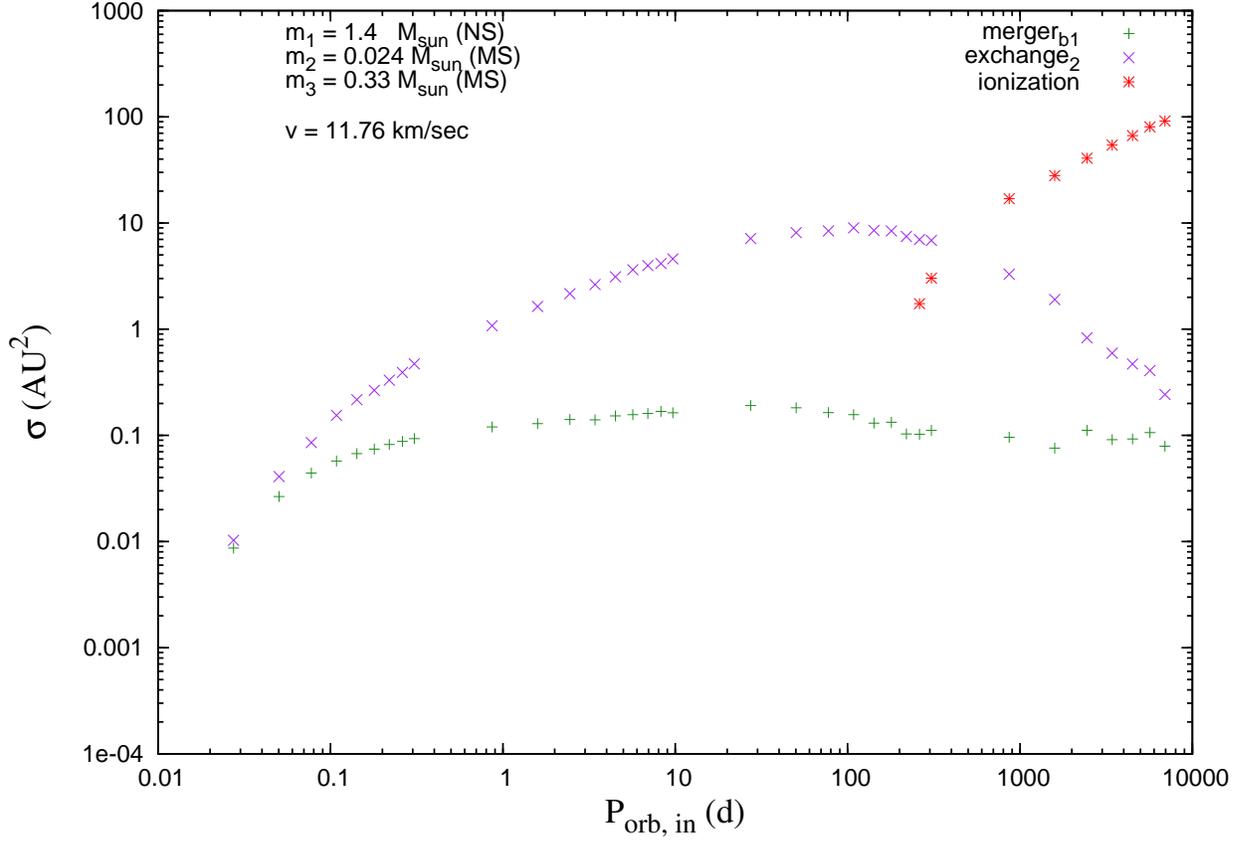} \caption{Variation of $\sigma$ for
exchange (purple `x'), merger (green `+') and ionization (red
`$\ast$') with $P_{orb,~in}$ for
$m_1=1.4~M_{\odot},~m_2=0.024~M_{\odot}, ~m_3=0.33~M_{\odot}$ and
$v=11.76$ km/sec. \label{fig:ion_lowmass}}
\end{figure}

STARLAB runs show that ionization starts only at high values of
$P_{orb}$ and a lower value of $m_2$ or a higher value of
$m_3$ can facilitate ionization for obvious reasons of available
kinetic and binding energies. As an example, keeping
$m_1=~1.4~M_{\odot}$ and $m_3=~0.33~M_{\odot}$ fixed, the minimum
value of $P_{orb}$ increases from $\sim 200$ days to $\sim 3000$
days when we increase $m_2$ from $0.024~M_{\odot}$ to
$0.16~M_{\odot}$. For the case of $m_2=~0.40~M_{\odot}$, ionization starts
after 10000 days. In Fig. \ref{fig:ion_lowmass}, we show the
variation of the cross section $\sigma$ with $P_{orb}$ for
different interaction with $m_1=~1.4~M_{\odot}$,
$m_2=~0.024~M_{\odot}$, $m_3=~0.33~M_{\odot}$; this is the only case
where ionization starts in the orbital period range of the observed
globular cluster binaries. This happens because the condition of
ionization $e.g.$ $v/v_c > 1$ is satisfied (where $v_c$ is the
critical velocity giving three unbound stars at zero energy
defined in Eq. \ref{eq:vc}). We have verified that
STARLAB outputs also match with
the analytical expression of scaled cross section $X$ given by
(Spitzer 1987) \
\begin{equation} X~=~\frac{\sigma}{\pi~a_{in}^2}\left(\frac{v}{v_c}
\right)^2 =\frac{40}{3}\frac{m_3^3}{m_1 m_2 (m_1+m_2+m_3)}
\end{equation}.

There are three disk pulsar
with $P_{orb}>1000$ days as J0823+0159 ($P_{orb} = 1232.40$ days,
$M_c = 0.23~M_{\odot}$), J1302-6350 ($P_{orb} = 1236.72$ days, $M_c = 4.14~M_{\odot}$) and
J1638-4725 ($P_{orb} = 1940.9$ days, $M_c = 5.84~M_{\odot}$). But such high
period pulsars have not been observed in globular clusters. If the evolution of
neutron star binaries are nearly the same in globular
 clusters as that in the galactic disk, there should have been similar long
period binaries with comparatively
 massive companions for which ionization cannot be very effective
as seen from our calculations.
It is possible that such systems have not yet been observed
because of observational selection effects, as long period searches require
correspondingly long time baselines.
Perhaps future pulsar searches with better sensitivity and longer time baselines
may reveal such binaries.

\section{Discussions}
\label{sec:disc}

Three globular cluster binary pulsars with $0.01< e <0.1$
and $60 < P_{orb} < 256 \; \rm d$ (see Fig 1) in M53, M3(D), and Ter 5(E) have companions of mass in the range $m_c = 0.21 - 0.35
\; \rm M_{\odot}$ ($i = 60^{\circ}$). These are possibly white
dwarf cores of red giant companions that overflowed Roche lobe
\cite{web83}. Such binaries would normally have the ``relic"
eccentricities $10^{-4}$ (green dashed lines). The above binary
pulsars with their presently mildly high eccentricities, may have
undergone fly-by encounters with field stars, rather than exchange or merger
interactions, which would have produced very high eccentricities
$e > 0.1$. Ter 5 E lies outside the core where $v_{10}/n_4$ is
higher than the central value, but even then, it might have
been be low enough to allow a strong fly-by interaction. It could
also have been ejected out of the high density core after a
strong interaction.

Another set of three globular cluster millisecond pulsars have
$0.01 < e < 0.1$, $2 < P_{orb} < 10 \; \rm d$; Ter 5 (W), 47 Tuc (H) and
NGC6440 (F). All these clusters have low values of $v_{10}/n_4$ (Table 2), and so
fly-by encounters in these clusters would be efficient
and could generate these eccentricities in GCs, even if their progenitor
binaries had shorter orbital periods and had sub-giant companions of the NSs
like the galactic equivalents of the red triangles of Fig 1. Alternately,
these binaries could also have been formed by fly-by interactions from a now
less abundant longer period $2 < P_{orb} < 10 $ day cluster of
``intermediate eccentricity" binaries to the right of the pulsars
(seen among the purple triangles in the middle of
Fig. \ref{fig:compare_gc_gal}).

The ``intermediate eccentricity" binary pulsars (group II : $0.01
> e \geq 2 \times 10^{-6}$), themselves could have been generated
by fly-by encounters with low (or ``zero") eccentricity
progenitor pulsars below the line of ``timing sensitivity limit"
(group III pulsars). A great majority of group II pulsars are
millisecond pulsars occurring in GCs with high probability of
fly-by encounters. But if some of the shorter $P_{orb}$ ($\sim 0.2$
day) pulsars have indeed been kicked-up to even higher
eccentricities in the past, their eccentricity would be reduced by
gravitational radiation again. The progenitor group III pulsars,
themselves occur in regions of favorable fly-by encounters
inducing higher eccentricities. These nearly circular binaries
have their $P_{orb}$ in the range of $0.106-4.0$ days. They can
not be results of exchange or merger as these processes produce
high eccentricities. Camilo \& Rasio (2005) discussed the
dynamical formation of ultra-compact binaries involving
intermediate mass main sequence stars in the early life of the GC
as the origin of the group III pulsars. These companions must
have been massive enough (beyond the present day cluster turn-off
mass of $0.8 \; M_{\odot}$) so that the initial mass transfer
became dynamically unstable, leading to common envelope evolution
and subsequent orbital decay and circularization. Alternately,
the present day red giant and NS collisions lead to a prompt
disruption of the red giant envelope and the system ends up as
eccentric NS-WD binary \cite{ras91}. These binaries could decay
to the group III pulsars by gravitational radiation if they had
short $P_{orb} \leq 0.2 \; \rm d$.

Despite the circumstance that most radio pulsars in GCs can be
explained by fly-by and/or exchange interactions, there exists a
pulsar like the PSR B1639+36B (i.e. M13B) which is a recycled
pulsar and in a region of the phase space that is difficult to
reach with either fly-by or exchange mechanisms. Systems like
this could have undergone eccentricity pumping of the inner orbit
due to the presence of a third star in a wide outer orbit, i.e. in
a hierarchical triplet like the PSR B1620-26 in M4 system \cite{sson93, ras95b, thor99, ford00}. Rasio $et~al.$ (1995) showed that only one such triple system can form in a globular cluster like M4, but in clusters with higher values of $n_4$ and/or binary fraction the number can be a few, though at present PSR B1620-26 is the only known triple system in globular clusters. A similar mechanism has been suggested recently \cite{cha08} for a galactic disk binary pulsar PSR J1903+0327, a radio pulsar with a
rotational period of 2.15 milliseconds in a highly eccentric (e =
0.44) 95-day orbit around a solar mass companion star. The formation scenarios for such an unusual pulsar both in and out of GCs, and in particular the observational consequences for further radio timing of PSR J1903+0327 in terms of eccentricity pumping taking account of relativistic precession of the inner orbit periastron, has
been investigated \cite{gop09}.

\section{Conclusion}
\label{sec:conclu}

We find that the presently observed orbital eccentricity and
period data of GC binary pulsars are largely explained by
numerical scattering experiments on stellar interaction scenarios
of fly-bys and exchanges with field stars. Binaries with $e >
0.1$ are most probably the result of exchange or merger events
whereas binaries with $0.01 > e > 0.00001 $ are products of
fly-by of single stars. A number of wide
orbit intermediate eccentricity pulsars seen in the galactic
field are absent in the GC sample because they have been kicked
up to relatively high eccentricities by passing stars in the
dense stellar environments in GCs. In some GCs such as Ter 5, the
stellar densities are so high, and the velocity dispersion so
modest that the interaction timescale for exchange and fly-by
interactions is relatively short. In such GCs a typical binary
system may undergo multiple interactions. If the original binary
contains a spun-up millisecond pulsar in a relatively ``soft" binary,
then the exchange
interaction may produce a single millisecond pulsar in the
cluster. This may explain the higher incidence of single ms PSRs
in GCs compared to that in the galactic disk \cite{cam05}. Terzan
5, for example, contains 16 single ms PSRs out of a total of 33
pulsars. In addition, exchange interactions, as we have seen, can
lead to highly eccentric orbits and the system can be ejected
from the cluster core. If the last encounter took place not too
long ago, the system can be at a relatively large offset from the
cluster core, albeit being still spatially co-located with the GC.

It is however clear that there are systems in GCs whose orbital
characteristics cannot be explained by purely episodic
interactions; an example of this kind is PSR B1620-26, which is a
hierarchical triplet, whose pulsar orbital eccentricity is
affected by the presence of a third star in a loosely bound
system. There are possibly other examples of a similar kind where
there is eccentricity pumping from the long term presence of
companions in a triplet system.

We have also considered the effects of collision induced
ionization on the present day distribution of orbital parameters
of radio pulsars in GCs. In the galactic disk we find that there
exist several pulsars with $P_{orb} > 1200 d$ (although some of
them have more massive companions than the pulsar binaries in the
GCs), while the similar wide orbit binaries are missing from the
GCs. While it is tempting to speculate that these are missing
from the present day GCs because they have been ionized in the
past, we find that the ionization probability becomes substantial
only in a restricted domain of masses of the companion and
incoming stars, and that too for essentially periods greater than
$P_{orb} > 1000 d$. Binary pulsars with very low companion masses
can be ionized easily. On the other hand the apparent
lack of the corresponding examples of galactic disk long orbital period
binaries surviving due to their massive companions in GCs could also be
due to observational selection effects.
While ionization interaction can explain
lack of binary pulsars with $P_{orb} > 1000 d$ in globular
clusters, this process is currently not important, but could have
played a role in the past for very long period binaries. On the
other hand there are somewhat wide binaries in the present day
GCs with moderately high eccentricities (e.g. $0.01< e <0.1$ and
$60 < P_{orb} < 256 \; \rm d$) which could have arisen out of
fly-by exchanges from progenitor binaries with ``relic"
eccentricities $e \sim 10^{-4}$. Many galactic disk binary pulsars are
seen in the $e-P_{orb}$ plane predicted by Phinney due to the
fluctuation dissipation of convective eddies and the resultant
orbital eccentricities that are induced. These pulsars are
missing from the GC sample. There is no reason not to expect
these systems to form in the GCs (although due to the lower
metallicity of the stellar companions in GCs, they are expected
to lead to somewhat less wide systems \cite{web83}). This can be
explained by the substantial probability of them being knocked
out of their original phase space due to flyby interaction in GCs.

\acknowledgments We thank Roger Blandford, Avinash Deshpande, Shri Kulkarni, 
Douglas Heggie, Michael Kramer, Ben Stappers, Duncan Lorimer, Paulo Freire, 
Maura McLaughlin, Andrew Fruchter, Jason Kalirai, Scott Ransom, Avi Loeb, 
Josh Grindlay, Daniel Fabrycky, Saul Rappaport, Victoria Kaspi, 
Patrick Lazarus, Fernando Camilo, Rino Bandiera, Sandro Mereghetti, 
Pierre Pizzochero and Masimo Turatto for discussions 
and Sayan Chakraborti for comments on the manuscript. We thank the STARLAB 
development group for the software and the ATNF pulsar group and 
Paulo Freire for pulsar
data bases.  We thank the participants and organizers of the
$2^{nd}$ IIA-PennState Astrostatistics School (Kavalur, 2008), in
particular Jogesh Babu, Sabyasachi Chatterjee and Prajval Shastri
for discussions. This manuscript was completed at Raman Research Institute
and we thank its Director, Ravi Subrahmanyan and its staff
for their hospitality on multiple occasions. This research 
is a part of 11th plan project 11P-409 at TIFR.

%%%%%%%%%%%%%%%%%%%%%%%%%%%%%%%%%

\clearpage

\clearpage

\clearpage

\clearpage

\clearpage

\clearpage

\begin{table*}
\caption{STARLAB runs with neutron star mass
$m_1~=~1.40~M_{\odot}$ and radius $R_1~=$ 10 km throughout for
$v=11.76$ km/sec.}
\begin{center}
\begin{tabular}{|l|l|c|c|c|}
\hline
Set & $m_1,~ m_2, ~m_3$ & $R_2, ~R_3$ & $v_f/v_c$ & $P_{orb,~in}$  \\
& $(M_{\odot})$ &  $(R_{\odot})$  & & (day) \\   \hline
1 & $1.40,~0.40,~0.33$ & $0.40~ {\rm (MS)},~ 0.33~{\rm (MS)}$  & $[0.018, 0.788]$ & $[0.07,6159]$  \\
2 & $1.40,~0.16,~0.33$ & $0.16~ {\rm (MS)},~0.33~{\rm (MS)}$  & $[0.019, 1.231]$ & $[0.03, 6616]$  \\
3 &$1.40,~0.024,~0.33$  & $0.024~ {\rm (MS)},~0.33~ {\rm (MS)}$  & $[0.035, 3.153]$ & $[0.01, 6924]$  \\
4 &$1.40,~0.40,~0.80$  & $0.40~ {\rm (MS)},~0.80~ {\rm (MS)}$  &  $[0.025, 1.110]$ & $[0.07, 6159]$  \\
5 & $1.40,~0.40,~0.80$ & $0.40~{\rm (MS)},~0.01~ {\rm (WD)}$  & $[0.025, 1.110]$  &$[0.07, 6159]$   \\
6 & $1.40,~0.16,~0.80$ & $0.16~{\rm (MS)},~0.80~ {\rm (MS)}$  & $[0.027, 1.716]$ &$[0.03, 6616]$   \\
7 & $1.40,~0.16,~0.80$ & $0.16~ {\rm (MS)},~0.01~ {\rm (WD)}$  & $[0.027, 1.716]$ & $[0.03, 6616]$    \\
8 & $1.40,~0.024,~0.80$  & $0.024~ {\rm (MS)},~0.80~ {\rm (MS)}$  &  $[0.049, 4.360]$ & $[0.01, 6924]$   \\
9 & $1.40,~0.024,~0.80$ & $0.024~ {\rm (MS)},~0.01~ {\rm (MS)}$  & $[0.049, 4.360]$ & $[0.01, 6924]$  \\
10 & $1.40,~0.40,~0.50$ & $0.40~ {\rm (MS)},~ 0.50~{\rm (MS)}$  & $[0.020,0.933]$ & $[0.07,6159]$  \\
11 & $1.40,~0.16,~0.50$ & $0.16~ {\rm (MS)},~0.50~{\rm (MS)}$  & $[0.023,1.452]$ & $[0.03,6616]$  \\
12 &$1.40,~0.024,~0.50$  & $0.024~ {\rm (MS)},~0.50~ {\rm (MS)}$  & $[0.041, 3.706,]$ & $[0.01,6924]$  \\
\hline
\end{tabular}
\end{center}
\label{tb:starlab_runs} \end{table*}

%%%%%%%%%%%%%%%%%%%%%%%%%%%%%%%%%%%%%%%

\clearpage

\begin{table}
\caption{Globular cluster parameters taken from Webbink (1985).}
\begin{center}
\begin{tabular}{cccccc}
\hline
group &GC & d (from sun) & $v_{10}$  &$n_{4}$ & $v_{10}/n_{4}$  \\
& &  (kpc) & ($10~ {\rm km /sec}$)  & ($10^4~\rm{pc^{-3}}$)  & \\ \hline
1&Ter 5 & 10.3 & 1.18  & 479.77 & 0.0024 \\ \hline
2&NGC 6440 & 8.4 & 1.30   & 102.10 & 0.013 \\
&M 30 & 8.0 & 0.52  & 36.39 & 0.014 \\
&NGC 1851  &12.1 &  0.98  &60.40  & 0.016  \\
&M 62 & 6.9 & 0.52  & 32.36 & 0.016 \\
&M 15 & 10.3  & 0.86  & 36.65  &  0.023 \\ \hline
3&NGC 6441 & 11.2 & 1.44  & 31.70 & 0.046 \\
&NGC 6544 & 2.7 & 0.59  & 9.75 & 0.060 \\
&47 Tuc & 4.5 & 1.32  & 31.99 & 0.062 \\ \hline
4&M 28 & 5.6 & 1.06 &  9.59 & 0.110 \\
&NGC 6342   & 8.6 & 0.45  &3.49 & 0.130 \\
&NGC 6752   & 4.0 & 0.78   & 5.97 & 0.131 \\
&NGC 6760   & 7.4 & 0.58  & 4.17 & 0.138 \\
&NGC 6539   & 8.4 & 0.45   & 3.09 & 0.146 \\
&NGC 6397   & 2.3 & 0.48  & 3.16 & 0.152 \\ \hline
5&M 4 &  2.2 & 0.51  &   1.63 & 0.32 \\
&M 5 & 7.5 & 0.84  &   2.57 & 0.328  \\
&M 3 & 10.4 &0.82  & 1.44 & 0.568  \\
&M 22 & 3.2 & 0.83 &  1.96 & 0.423 \\ \hline
6&M 71 & 4.0 & 0.33 &  0.35 & 0.934 \\
&M 13 & 30.4 & 0.78 &  0.72 & 1.082 \\
&NGC 6749   & 7.9 &  0.39    & 0.23 & 1.713\\
&M 53 & 17.8 & 0.65  & 0.30 & 2.167 \\
\hline
\end{tabular}
\end{center}
\label{tb:gc_v_n}
\end{table}

\begin{table*}
\caption{Parameters for 73 binary pulsars with known orbital
solutions in 23 globular clusters. Pulsar offsets from the
cluster cores are in the units of the core radius $r_c$ (taken
from Freire's webpage at
http://www.naic.edu/$\sim$pfreire/GCpsr.html and S. Ransom's
webpage at http://www.cv.nrao.edu/$\sim$sransom/). In Freire's webpage, eccentricities of 32 pulsars are listed as zero, which we have replaced by an arbitrarily small value 0.0000003 (see text for details). Last column shows the median values of companion masses $i.e.$ for inclination angle $i=60$ degree (again from Freire's webpage) taking pulsar masses as $1.35~ M_{\odot}$.  For some pulsars, better mass measurements are available from the measurement of rates of periastron advance or from the spectral analysis of the optical counterpart in case of PSR J1911-5958A. These pulsars are listed here : (i) 47Tuc H has $m_p = 1.44~M_{\odot},~ m_c = 0.17~M_{\odot}$ (Freire $et~al.$ 2003) (ii) NGC 1851A has $m_p = 1.35~M_{\odot},~ m_c = 1.103~M_{\odot}$ (Freire, Ransom \& Gupta 2007) (iii) M5B has $m_p = 2.08~M_{\odot},~ m_c = 0.21~M_{\odot}$ (Freire, Wolszczan, van den Berg \& Hessels, 2008) (iv) Ter 5 I has $m_p = 1.87~M_{\odot},~ m_c = 0.30~M_{\odot}$ (Ransom $et~ al.$ 2005) (v) Ter 5 J has $m_p = 1.73~M_{\odot},~ m_c = 0.47~M_{\odot}$ (Ransom $et~ al.$ 2005) (vi) NGC 6440B has $m_p = 2.74~M_{\odot},~ m_c = 0.18~M_{\odot}$ (Freire $et~ al.$ 2008) (vii) NGC 6441A has $m_p = 1.26~M_{\odot},~ m_c = 0.67~M_{\odot}$ (Freire $et~ al.$ 2008) (viii) PSR J1911-5958A has $m_p = 1.34~M_{\odot},~ m_c = 0.175~M_{\odot}$ (Bassa, van Kerkwijk, Koester \& Verbunt 2006) (using spectral analysis of the optical counterpart ) (ix) M15 C has $m_p = 1.358~M_{\odot},~ m_c = 1.354~M_{\odot}$ (Jacoby $et~al.$ 2006). The eclipsing pulsars are marked with ``(e)" which most probably have main sequence companions (unless the inclination angle is very close to $90^{\circ}$).} \footnotesize
\label{tab:psr_parms}
\begin{center}
\begin{tabular}{llllllllll}
\hline \hline \multicolumn{1}{l}{No}          &
\multicolumn{1}{l}{GC}             &
\multicolumn{1}{l}{Pulsar}        &
\multicolumn{1}{l}{offset}             &
\multicolumn{1}{l}{$P_{s}$}           &
\multicolumn{1}{l}{$\dot{P_{s}}$}            &
\multicolumn{1}{l}{DM} &
\multicolumn{1}{l}{$P_{orb}$}             &
\multicolumn{1}{l}{$e$}             &
\multicolumn{1}{l}{$m_c$}             \\
\multicolumn{1}{l}{}            & \multicolumn{1}{l}{}    &
\multicolumn{1}{l}{}         & \multicolumn{1}{l}{(in} &
\multicolumn{1}{l}{}             &
\multicolumn{1}{l}{$(10^{-20}$}          &
\multicolumn{1}{l}{} & \multicolumn{1}{l}{} &
\multicolumn{1}{l}{} & \multicolumn{1}{l}{} \\
\multicolumn{1}{l}{}            & \multicolumn{1}{l}{}    &
\multicolumn{1}{l}{}         &
\multicolumn{1}{l}{$r_c$)}             &
\multicolumn{1}{l}{(ms)}             & \multicolumn{1}{l}{${\rm
sec/sec})$}         & \multicolumn{1}{l}{${(\rm cm^{-3} pc})$} &
\multicolumn{1}{l}{(d)} & \multicolumn{1}{l}{} &
 \multicolumn{1}{l}{($M_\odot$)}
 \vspace{1mm}  \\ \hline \hline
1 & 47~Tuc & J0024-7205E &  1.477 &   3.536  & 9.851 & 24.23 &  2.25684 &     0.0003152    &    0.18  \\
2 & 47~Tuc & J0024-7204H &  1.75  &   3.210  & -0.183    & 24.36 &  2.35770 &   0.070560     &  0.19     \\
3 & 47~Tuc &J0024-7204I &  0.659 &   3.485  &  -4.587      &  24.42 &  0.22979 &   $<0.0004$        &  0.015   \\
4 & 47~Tuc &J0023-7203J (e) &  2.273 &   2.100  & -0.979     & 24.58 & 0.12066 &      $<0.00004$  &       0.024   \\
5 &  47~Tuc &J0024-7204O &  0.136 &   2.643  & 3.035       & 24.36 &  0.13597  &    $<0.00016$    &    0.025   \\
6 &  47~Tuc &J0024-7204P  &  ?     &  3.643  & ? & 24.30 & 0.1472   &    0.0000003 &     0.02   \\
7 &  47~Tuc &  J0024-7204Q &  2.227 &   4.033 & 3.402    &  24.29&  1.18908 &   0.000085    &    0.21   \\
8 &  47~Tuc & J0024-7204R (e) &  ?     &  3.480   & ? & 24.40 & 0.0662    &  0.0000003 &     0.030  \\
9  &  47~Tuc & J0024-7204S &  0.432  &  2.830  & -12.054  & 24.35 &  1.20172  &  0.000394   &    0.10  \\
10 &  47~Tuc & J0024-7204T &  0.773  &  7.588  & 29.37    &  24.39 &  1.12618  &  0.00040    &    0.20  \\
11  & 47~Tuc & J0024-7203U &  2.136  &  4.343  & 9.523    & 24.33 &  0.42911  &  0.000149   &    0.14  \\
12 &  47~Tuc & J0024-7204V  (e?) &   ?    &  4.810    & ? & 24.10 & 0.227   & 0.0000003    &   0.35  \\
13  &  47~Tuc & J0024-7204W (e) & 0.182  &  2.352  & ? & 24.30 &  0.1330  &  0.0000003  &    0.14    \\
14  & 47~Tuc &  J0024-7204Y  &   ?    &  2.197  & ? & 24.20 & 0.52194  &     0.0000003   &    0.16  \\
\hline
15 & NGC1851 & J0514-4002A  & 1.333  &  4.990  & 0.117    & 52.15 & 18.78518 &   0.8879773    &    1.10 \\
\hline
16 & M53 & B1310+18    &     ?   &   33.163    & ? & 24.00 & 255.8    &     0.01  &  0.35  \\
\hline
17  & M3 &    J1342+2822B &  0.254   &  2.389   & 1.858   & 26.15 &  1.41735  &   0.0000003   &    0.21     \\
18 & M3  &   J1342+2822D &  0.418   &  5.443  & ? & 26.34 &
128.752 & 0.0753      &   0.21   \\
\hline
\hline
\end{tabular}
\end{center}
\end{table*}

\addtocounter{table}{-1}
\begin{table*}
\caption{(continued).}
\vspace{-2.0mm}
\footnotesize
\begin{center}
\begin{tabular}{llllllllll}
\hline \hline \multicolumn{1}{l}{No}          &
\multicolumn{1}{l}{GC}             &
\multicolumn{1}{l}{Pulsar}        &
\multicolumn{1}{l}{offset}             &
\multicolumn{1}{l}{$P_{s}$}           &
\multicolumn{1}{l}{$\dot{P_{s}}$}            &
\multicolumn{1}{l}{DM} &
\multicolumn{1}{l}{$P_{orb}$}             &
\multicolumn{1}{l}{$e$}             &
\multicolumn{1}{l}{$m_c$}             \\
\multicolumn{1}{l}{}            & \multicolumn{1}{l}{}    &
\multicolumn{1}{l}{}         & \multicolumn{1}{l}{(in} &
\multicolumn{1}{l}{}             &
\multicolumn{1}{l}{$(10^{-20}$}         &
\multicolumn{1}{l}{} & \multicolumn{1}{l}{} &
\multicolumn{1}{l}{} &
\multicolumn{1}{l}{}\\
\multicolumn{1}{l}{}            & \multicolumn{1}{l}{}    &
\multicolumn{1}{l}{}         &
\multicolumn{1}{l}{$r_c$)}             &
\multicolumn{1}{l}{(ms)}             & \multicolumn{1}{l}{${\rm
sec/sec})$}         & \multicolumn{1}{l}{${(\rm cm^{-3} pc})$} &
\multicolumn{1}{l}{(d)} & \multicolumn{1}{l}{} &
\multicolumn{1}{l}{($M_\odot$)}
\vspace{1mm}  \\
\hline  \hline
19 & M5 &     B1516+02B  &   0.545  &   7.947  & -0.331      & 29.45 & 6.85845  &   0.13784    &   0.13   \\
20 & M5 &   J1518+0204C (e)  &  ?   &    2.484   & ? & 29.30 & 0.087     &  0.0000003    &   0.038   \\
21 & M5 &  J1518+0204D  &   ?   &    2.988  &  ?  & 29.30 & 1.22
& 0.0000003     &   0.20
  \\
22 & M5 &   J1518+0204E  &   ?   &    3.182   & ? & 29.30  & 1.10  &  0.0000003   &    0.15  \\
\hline
23 & M4 & B1620-26   &   0.924  &  11.076  & -5.469      & 62.86 & 191.44281  &  0.02531545     &   0.33   \\
\hline
24 & M13 & B1639+36B     &  ?  &   3.528  & ? & 29.50 & 1.25911  &  $<0.001$   &   0.19   \\
25  & M13 &  J1641+3627D  &   ?   &     3.118   & ? & 30.60 & 0.591     & 0.0000003    &   0.18    \\
26  & M13 &   J1641+3627E (e?) &    ?  &      2.487  & ? & 30.30 &  0.117   &   0.0000003     &   0.02  \\ \hline
27 & M62 &    J1701-3006A  & 1.778  &    5.241 & -13.196       & 115.03  & 3.80595 & 0.000004    &    0.23  \\
28 & M62 &  J1701-3006B (e) & 0.155    &  3.594  & -34.978       & 113.44 & 0.14455  & $<0.00007$       &  0.14   \\
29 & M62 &   J1701-3006C &  0.972 &     3.806 &-3.189       & 114.56 & 0.21500 & $<0.00006$     &   0.08   \\
30  & M62 &  J1701-3006D  &   ?      &  3.418    & ? & 114.31 & 1.12     & 0.0000003     &   0.14  \\
31  & M62 &   J1701-3006E (e) &   ?     &   3.234   & ? & 113.78  & 0.16   &    0.0000003    &   0.035   \\
32  & M62 &   J1701-3006F  &   ?    &    2.295  & ? & 113.36 & 0.20 &   0.0000003     &   0.02  \\
\hline
33 & NGC6342  & B1718-19 (e)  &  46.000 &  1004.04   & $1.59 \times 10^{5}$  & 71.00 & 0.25827 &  $<0.005$     &   0.13    \\ \hline
34 & NGC6397 & J1740-5340 (e) &  18.340   &   3.650  & 16.8    & 71.80 & 1.35406 &   $<0.0001$    &     0.22    \\
\hline
35 & Ter5  &  J1748-2446A (e) &  2.778  &  11.5632 & -3.400     & 242.10 & 0.075646 &  0.0000003    & 0.10    \\
36 & Ter5  &  J1748-2446E   & out(1.6)  & 2.19780 & ?     & 236.84 & 60.06   &    0.02    &        0.25    \\
37 & Ter5  &  J1748-2446I   &  ?    &     9.57019  & ?    & 238.73 & 1.328 &     0.428   &   0.24   \\
38 & Ter5  &  J1748-2446J   &  ?    &    80.3379  & ?    & 234.35 & 1.102 &     0.350  &        0.39   \\
39 & Ter5  &  J1748-2446M  & in(0.48) &  3.56957 & ?    & 238.65 & 0.4431 & 0.0000003        & 0.16    \\
40 & Ter5 &   J1748-2446N  & in(0.39) &  8.66690 & ? &    238.47 & 0.3855 & 0.000045   &    0.56       \\
41 & Ter5 &   J1748-2446O (e) &  in(0.45) &  1.67663 & ? & 236.38 & 0.2595  &   0.0000003    &   0.04  \\
42 & Ter5  &  J1748-2446P (e) &  in(0.74)  & 1.72862 & ? & 238.79 & 0.3626  &   0.0000003    &   0.44   \\
43 & Ter5  &  J1748-2446Q  & out(1.45) &  2.812  & ? & 234.50 & 30.295   &  0.722   &   0.53      \\
44 & Ter5 &   J1748-2446U   &   ?     &   3.289   & ? & 235.46 & 3.57   &     0.61 &         0.46    \\
45 & Ter5  &  J1748-2446V &  in(0.90)  & 2.07251  & ? & 239.11  & 0.5036  &  0.0000003    &   0.14        \\
46 & Ter5  &  J1748-2446W &  in(0.42)  & 4.20518 &  ? & 239.14 & 4.877 &  0.015   &    0.34     \\
47 & Ter5   &  J1748-2446X    &  ?    &    2.99926  & ? & 240.03 & 4.99850 &  0.3024     &    0.29        \\
48 & Ter5 &   J1748-2446Y  &  in(0.55)  & 2.04816 &  ? & 239.11 & 1.16443  & 0.00002     &    0.16       \\
49 & Ter5 &   J1748-2446Z   &   ?   &     2.46259  & ? & 238.85 & 3.48807     & 0.7608     &   0.25      \\
50 &Ter5  &  J1748-2446ad (e) &   ?   &     1.39595  & ? & 235.60 & 1.09443     &  0.0000003      &  0.16     \\
51 & Ter5 &   J1748-2446ae &  in(0.42)   & 3.65859  & ? & 238.75
& 0.17073 &  0.0000003       & 0.019 \\  \hline
52 & NGC6440 & J1748-2021B  & 0.530   &   16.760 & -32.913 & 220.92 & 20.550 &     0.570     &   0.090  \\
53  & NGC6440  & J1748-2021D (e) &  4.230    &   13.496 &  58.678           & 224.98 &  0.286 & 0.0000003       & 0.14  \\
54  & NGC6440  &  J1748-2021F   & 0.690    &   3.794 & 31.240     & 224.10 &  9.83397  &  0.0531    &   0.35 \\
\hline
55 & NGC6441 & J1750-37A   &  1.910   &  111.609  & 566.1 &     233.82 &17.3   &     0.71   &   0.70    \\
56 & NGC6441 & J1750-3703B &  3.000   &    6.074    & 1.92 &      234.39 & 3.61   & 0.0000003      &   0.19   \\ \hline
 57 & NGC6539 & B1802-07   &   0.463   &   23.1009  & 47.0 &     186.38 &  2.61676  &   0.21206     &    0.35    \\ \hline
58 & NGC6544 & B1802-07    &   ?     &     3.05945  & ? & 134.0 & 0.071092   & 0.0000003       & 0.010   \\ \hline
\hline
\end{tabular}
\end{center}
\end{table*}

\addtocounter{table}{-1}
\begin{table*}
\caption{(continued).}
\vspace{-2.0mm}
\footnotesize
\begin{center}
\begin{tabular}{llllllllll}
\hline \hline \multicolumn{1}{l}{No}          &
\multicolumn{1}{l}{GC}             &
\multicolumn{1}{l}{Pulsar}        &
\multicolumn{1}{l}{offset}             &
\multicolumn{1}{l}{$P_{s}$}           &
\multicolumn{1}{l}{$\dot{P_{s}}$}            &
\multicolumn{1}{l}{DM} &
\multicolumn{1}{l}{$P_{orb}$}             &
\multicolumn{1}{l}{$e$}             &
\multicolumn{1}{l}{$m_c$}             \\
\multicolumn{1}{l}{}            & \multicolumn{1}{l}{}    &
\multicolumn{1}{l}{}         & \multicolumn{1}{l}{(in} &
\multicolumn{1}{l}{}             &
\multicolumn{1}{l}{$(10^{-20}$}         &
\multicolumn{1}{l}{} & \multicolumn{1}{l}{} &
\multicolumn{1}{l}{} &
\multicolumn{1}{l}{}\\
\multicolumn{1}{l}{}            & \multicolumn{1}{l}{}    &
\multicolumn{1}{l}{}         &
\multicolumn{1}{l}{$r_c$)}             &
\multicolumn{1}{l}{(ms)}             & \multicolumn{1}{l}{${\rm
sec/sec})$}         & \multicolumn{1}{l}{${(\rm cm^{-3} pc})$} &
\multicolumn{1}{l}{(d)} & \multicolumn{1}{l}{} &
\multicolumn{1}{l}{($M_\odot$)}
\vspace{1mm}  \\
\hline  \hline
59 & M28  &   J1824-2452C  &  ?   &     4.159    & ? & 120.70 & 8.078    &  0.847    &      0.30     \\
60 & M28  &   J1824-2452D &   ?    &   79.832  & ? & 119.50 & 30.404 &   0.776  &      0.45     \\
61 & M28  &   J1824-2452G  &  ?   &     5.909   & ? & 119.40 & 0.1046 &  0.0000003     &   0.011    \\
62 & M28  &   J1824-2452H (e) &  ?   &    4.629   & ? & 121.50 & 0.435     & 0.0000003     &  0.20     \\
63 & M28  &   J1824-2452I (e) & ?    &    3.93185 & ? & 119.00 & 0.45941 &  0.0000003 &     0.20     \\
64 & M28  &   J1824-2452J   &  ?   &    4.039    & ? & 119.20 & 0.0974    &  0.0000003    &    0.015    \\
65 & M28  &   J1824-2452K  &  ?    &    4.46105 & ? & 119.80 & 3.91034  &  0.001524      &    0.16     \\
66 & M28   &  J1824-2452L   & ?    &    4.10011  & ? & 119.00 &0.22571  &   0.0000003     &  0.022    \\ \hline
67 & M22   &  J1836-2354A  &  ?   &       3.35434  & ? & 89.10 & 0.20276   &  0.0000003       &   0.020    \\ \hline
68  & NGC6749  & J1905+0154A   &  0.662    &    3.193  & ? &  193.69 &   0.81255  &  0.0000003     &    0.090    \\ \hline
69  & NGC6752  &  J1911-5958A   & 37.588    &   3.26619 & 0.307 &       33.68 &  0.83711  &  $<0.00001$      &   0.22     \\ \hline
70  & NGC6760  & J1911+0102A   &  1.273   &    3.61852 & -0.658  & 202.68 &  0.140996  &   $<0.00013$       &   0.020      \\ \hline
71  & M71   &   J1953+1846A (e)   &  ?      &    4.888   & ? & 117.00 & 0.1766  &  0.0000003     &    0.032     \\ \hline
72  & M15    &  B2127+11C    &  13.486   &   30.5293   & 499.1      & 67.13 & 0.33528   &    0.681386     &    1.13     \\ \hline
73  & M30    &  J2140-2310A (e)  &  1.117    &  11.0193   & -5.181      & 25.06 & 0.17399   &    $<0.00012$       &   0.11     \\ \hline
\hline

\hline
\end{tabular}
\end{center}
\end{table*}

\end{document}